\ifx\onecol\undefined
\documentclass[journal,twocolumn]{IEEEtran}
\else 
\documentclass[12pt,draftclsnofoot,journal,onecolumn]{IEEEtran}
\fi

\usepackage{amsfonts}
\usepackage{amsmath,amssymb}
\usepackage{acronym}  
\usepackage{algorithm}
\usepackage{algorithmic}
\usepackage{bm}
\usepackage{bbm}
\usepackage{booktabs}
\usepackage{color, soul}
\usepackage{cite}
\usepackage{graphicx}
\usepackage{indentfirst}
\usepackage{multirow}
\usepackage{setspace}
\usepackage{tikz}
\usepackage{threeparttable}
\usetikzlibrary{arrows}
\usepackage{subfigure}
\usepackage[amsmath,thmmarks]{ntheorem}
\usepackage{theorem}
\usepackage{hyperref}
\hypersetup{hidelinks, 
colorlinks=true,
allcolors=black,
pdfstartview=Fit,
breaklinks=true}

\newtheorem{theorem}{\bf Theorem}

\newtheorem{lemma}{\bf Lemma}

\newtheorem{remark}{\bf Remark}

\theoremheaderfont{~~~\it}\theorembodyfont{\upshape}%
\theoremstyle{nonumberplain}
\theoremseparator{}
\theoremsymbol{\rule{1ex}{1ex}}

\acrodef{SNR}[SNR]{signal-to-noise ratio}
\acrodef{DoF}[DoF]{degree of freedom}
\acrodef{FPGA}[FPGA]{field programmable gate array}
\acrodef{RF}[RF]{radio-frequency}
\acrodef{BS}[BS]{base station}
\acrodef{DL}[DL]{downlink}
\acrodef{TA}[TA]{transmit antenna}
\acrodef{RA}[RA]{receive antenna}
\acrodef{LOS}[LoS]{line-of-sight}
\acrodef{NLOS}[NLoS]{none-line-of-sight}
\acrodef{RC}[RC]{receiver combining}
\acrodef{AWGN}[AWGN]{additive white Gaussian noise}
\acrodef{DFT}[DFT]{discrete Fourier transform}
\acrodef{IRF}[IRF]{interference random field}
\acrodef{MISO}[MISO]{multiple-input single-output}
\acrodef{CSI}[CSI]{channel state information}
\acrodef{CRLB}[CRLB]{Cram{\'e}r-Rao lower bound}
\acrodef{UPA}[UPA]{uniform planar array}

\def \T {\bm \Theta}

\def \diag {\text{diag}}

\def \exp {\text{exp}}
\def \arg {\text{arg}}
\def \CN {\mathcal{CN}}
\def \VM {\mathcal{VM}}
\def \re {\text{Re}}
\def \nc {\mathcal{NC}}

\def \T {^{\mathsf{T}}}
\def \H {^{\mathsf{H}}}
\def \ri {{\rm i}}
\def \d {{\rm d}}

\newcommand{\RNum}[1]{\uppercase\expandafter{\romannumeral #1\relax}}
\def\AutoColumn#1#2{{\ifx\onecol\undefined {#2} \else {#1} \fi}}

\ifx\onecol\undefined
\newcommand{\myincludegraphics}[2][width=1\linewidth]{\includegraphics[#1]{#2}}
\else 
\newcommand{\myincludegraphics}[2][width=0.8\linewidth]{\includegraphics[#1]{#2}}
\fi
\newcommand{\red}[1]{{\color{black}{#1}}}

\ifCLASSINFOpdf
\else
\fi
\begin{document}
\title{Sensing RISs: Enabling Dimension-Independent CSI Acquisition for Beamforming}
\author{{Jieao~Zhu, Kunzan~Liu, Zhongzhichao~Wan, Linglong~Dai, Tie Jun Cui, and H.~Vincent~Poor }

\thanks{This work was supported in part by the National Key Research and Development Program of China (Grant No.2020YFB1807201), in part by the National Natural Science Foundation of China (Grant No. 62031019), in part by the U.S National Science Foundation under Grants CCF-1908308 and CNS-2128448, and in part by the National Natural Science Foundation of China under Grant 62288101. {\it (Corresponding author: Linglong Dai.)}}
\thanks{J. Zhu, K. Liu, Z. Wan, and L. Dai are with the Department of Electronic Engineering, Tsinghua University, Beijing 100084, China as well as the Beijing National Research Center for Information Science and Technology (BNRist) (e-mails: \{zja21, lkz18, wzzc20\}@mails.tsinghua.edu.cn, daill@tsinghua.edu.cn).}
\thanks{T. J. Cui is with the State Key Laboratory of Millimeter Waves, Southeast University, China (e-mail: tjcui@seu.edu.cn).}
\thanks{H. V. Poor is with the Department of Electrical and Computer Engineering, Princeton University, USA (e-mail: poor@princeton.edu).}

}

\markboth{IEEE Transactions on Information Theory}%
{Zhu \MakeLowercase{\textit{et al.}}: Sensing RISs: Enabling Dimension-Independent CSI Acquisition for Beamforming}

\maketitle

\begin{abstract}
Reconfigurable intelligent surfaces (RISs) are envisioned as a potentially transformative technology for future wireless communications.
\red{However, RISs' inability to process signals and the attendant increased channel dimension have brought new challenges to RIS-assisted systems, including significantly increased pilot overhead required for channel estimation. } To address these problems, several prior contributions that enhance the hardware architecture of RISs or develop algorithms to exploit the channels' mathematical properties have been made, where the required pilot overhead is reduced to be proportional to the number of RIS elements.
In this paper, we propose a dimension-independent channel state information (CSI) acquisition approach in which the required pilot overhead is independent of the number of RIS elements.
Specifically, in contrast to traditional signal transmission methods, where signals from the base station (BS) and the users are transmitted in different time slots, we propose a novel method in which signals are transmitted from the BS and the user simultaneously during CSI acquisition. 
\red{With this method, an electromagnetic interference random field (IRF) will be induced on the RIS, and we propose the structure of sensing RIS to capture its features. 
Moreover, we develop three algorithms for parameter estimation in this system, in which one of the proposed vM-EM algorithm is analyzed with the fixed-point perturbation method to obtain an asymptotic achievable bound. } 
In addition, we also derive the Cram{\'e}r-Rao lower bound (CRLB) and an asymptotic expression for characterizing the best possible performance of the proposed algorithms. 
Simulation results verify that our proposed signal transmission method and the corresponding algorithms can achieve dimension-independent CSI acquisition for beamforming. 
\end{abstract}

\begin{IEEEkeywords}
Reconfigurable intelligent surface (RIS), channel estimation, interference random field (IRF), dimension-independent CSI acquisition. 
\end{IEEEkeywords}

\section{Introduction}
Reconfigurable intelligent surfaces (RISs) are considered to be a potentially important technology for future wireless communications.
The characteristics of low cost and power consumption make RISs a promising solution for overcoming blockages, improving capacity, and reducing transmit power~\cite{basar2019wireless,liu2021compact,wu2019intelligent}.
Specifically, an RIS is a large-scale array composed of passive elements, which can achieve significant beamforming gain by appropriately imposing phase shifts on the incident electromagnetic waves~\cite{di2020smart}.
To achieve this beamforming gain, accurate \ac{CSI} should be acquired beforehand, which makes channel estimation an essential prerequisite for RIS-assisted communications~\cite{wei2021channel}.

Although channel estimation has been well investigated in conventional communication systems, the additional employment of RISs brings about two challenges~\cite{weili2021channel}.
Firstly, in contrast to traditional antenna array capable of transmitting, receiving, and processing the signals, RISs can only passively reflect the incident signals.
Secondly, since the number of RIS elements is usually large, the dimensions of the channels increase sharply compared with conventional communication systems, which results in unacceptably high pilot overhead for channel estimation.
These two main challenges result in the need for new channel estimation techniques for beamforming in RIS-assisted communications.

\subsection{Prior Works}
\label{Prior Works}
Generally, for RIS-assisted systems, channel estimation and beamforming are two separate procedures.
Channel estimation is performed first, and then the obtained \ac{CSI} is utilized for beamforming.
The beamforming gain relies heavily on the channel estimation accuracy.

To tackle the challenges mentioned above in channel estimation for RIS-assisted communications, some solutions have been proposed, which can be generally divided into two categories.
The first category modifies the hardware architecture of the RIS, which enables some signal processing capability.
For example, by sparsely replacing some of the RIS elements with active sensors capable of baseband processing, the authors of \cite{taha2021enabling} proposed a compressive sensing and deep learning-based channel estimation scheme with negligible pilot overhead. 
To further reduce the number of active sensors, in \cite{alexandropoulos2020hardware} the authors proposed an alternating direction method-based channel estimation procedure with a single \ac{RF} chain, with the help of an extended analog combiner~\cite{vlachos2019wideband}. Since in the above works, additional RF chains have already been attached to RISs for channel sensing, an analogous improvement is that these RF chains can also perform signal relaying. In~\cite{nguyen2021hybrid}, a hybrid relay-reflecting architecture is considered, where a few passive RIS elements are replaced by active amplify-and-forward relay modules. These active modules are then capable of channel estimation. 

The second category of channel estimation methods preserve the original hardware architecture of RISs, but they introduce algorithms to exploit the new structural channel properties that RISs bring about. 
Exploiting the two-timescale channel property in the RIS-assisted system, the authors of~\cite{Huchen} proposed a two-timescale channel estimation algorithm that reduces the pilot overhead in the time-averaged sense.
In other works~\cite{wang2020compressed,wei2021channel}, by exploiting the sparsity of channels in the angular domain, compressive sensing-based algorithms are developed with reduced pilot overhead.
In \cite{wang2020channel}, based on the shared reflected channels among multiple users, the authors proposed a three-phase channel estimation framework to further reduce the pilot overhead. 

Note that, in order to achieve high beamforming gain, all the above channel estimation methods are designed to estimate the full channel matrix as accurately as possible. However, since the channel matrix is at least of size $\mathcal{O}(N)$, where $N$ denotes the number of RIS elements, the required pilot overhead for channel estimation is usually proportional to the number of RIS elements~\cite{taha2021enabling,alexandropoulos2020hardware,vlachos2019wideband,nguyen2021hybrid,Huchen,wang2020compressed,wang2020channel,kundu2021channel} in practical systems, which makes most of the existing channel estimation schemes dimension-dependent. This is unacceptable especially when the RIS is fabricated with a large number of elements (e.g. 1100 elements in~\cite{pei2021ris}, 2304 elements in~\cite{yangfan2020coding}, and 10240 elements in~\cite{yang2018reconfigurable}).   
Therefore, the following question naturally arises: {\it Does there exist a dimension-independent approach where the required pilot overhead is independent of the number of RIS elements?}


\subsection{Our Contributions}
\label{Our Contributions}

We point out that the main drawback of conventional channel estimation methods is that they only extract the mathematical features of channels, while neglecting their electromagnetic nature. This limitation has caused the problem of high pilot overhead required for channel estimation in RIS-assisted systems. Thus, we propose an \ac{IRF}-based approach, where the \ac{IRF} induced on an RIS is utilized for channel estimation\footnote{Simulation codes are provided to reproduce the results in this paper: \url{http://oa.ee.tsinghua.edu.cn/dailinglong/publications/publications.html}.}.
Specifically, the contributions of this paper can be summarized as follows.
    \begin{itemize}
        \item 
        Inspired by optical interference where phase information can be obtained from the interference fringes, we reveal that phase information about the channel can also be gathered from the phenomenon of electromagnetic interference that occurs on RISs, which we name the \ac{IRF}. 
        To induce the \ac{IRF} on an RIS, we propose a novel pilot transmission method called simultaneous rotational signaling, where signals are transmitted from the \ac{BS} and the user to the RIS simultaneously, and the two signals bear a slight frequency difference during CSI acquisition. 
        \item
        To exploit this \ac{IRF} for CSI acquisition, we employ a  sensing RIS that integrates power detectors into the RIS elements to capture the features of the \ac{IRF}. Each of the power detectors can acquire its phase information independently from the \ac{IRF}, so that the required pilot overhead is independent of the number of RIS elements. 
        We then develop discrete Fourier transform (DFT), maximum likelihood (ML), and von Mises-expectation maximization (vM-EM) phase estimation algorithms to extract the phase information from the power detector signals in order to perform beamforming. 
        \item \red{
        By the fixed-point perturbation method, we analytically prove that the proposed vM-EM algorithm achieves an error decay of $\mathcal{O}(\bar{\gamma}^{-1})$, which is the best expectable asymptotic precision that a phase estimator could attain. Furthermore, we derive the \ac{CRLB} of the IRF phase estimation problem as well as its approximated asymptotic expression. 
        Our numerical results verify that the developed vM-EM algorithm achieves the theoretical error decay rate, and that it is close to the \ac{CRLB}.  
        Our simulation results further demonstrate that the proposed signal transmission method and the corresponding algorithms can realize dimension-independent CSI acquisition for beamforming, and can achieve near-optimal system spectral efficiency. }
    \end{itemize}

\subsection{Organization and Notation}

\textit{Organization:}
The rest of the paper is organized as follows.
In Section~\ref{System Model}, we introduce the system model of RIS-aided communications, and review the existing separate procedures of beamforming and channel estimation.
In Section~\ref{Interference Random Field}, we propose a novel signal transmission method for CSI acquisition in RIS-aided system, and reveal the \ac{IRF} phenomenon under this method.
In Section~\ref{Sensing RIS-Based Channel Estimation}, we introduce the hardware architecture of a sensing RIS to exploit the \ac{IRF}.
Based on the sensing RIS, we propose three algorithms to realize channel estimation.
In Section~\ref{Performance Analysis}, we first analyze the asymptotic performance of the proposed vM-EM algorithm, and then derive the \ac{CRLB} as well as its asymptotic expression for channel estimation in the sensing RIS-assisted system.
In Section~\ref{Simulation Results}, simulation results are provided for quantifying the performance of our proposed sensing RIS-based channel estimation as a novel dimension-independent solution to the RIS channel estimation problem.
Finally, in Section~\ref{Conclusion}, we provide our conclusions followed by promising future research ideas.

\textit{Notation:} $\mathbb C$ and $\mathbb R$ denote the set of complex and real numbers, respectively;
$\ri$ denotes the imaginary unit; 
$\{L\}$ represents the set of integers $\{0,1,\cdots,L-1\}$;
$\bm A^{-1}, \bm A^*,\bm A\T,$ and $\bm A\H$ denote the inverse, conjugate, transpose, and conjugate transpose of matrix $\bm A$, respectively; 
$\Vert \cdot\Vert_{2}$ is the $\mathcal{L}_{2}$-norm of its argument function;
$\Vert \cdot \Vert_{F}$ denotes the Frobenius norm of its argument matrix;  
$\langle{\bm x}, {\bm y}\rangle := {\bm x}\H {\bm y}$ denotes the inner product of complex vectors ${\bm x}, {\bm y}$; 
$\text{arg}(x)$ and $\text{exp}(x)$ denote the phase angle and exponential of the complex scalar $x$, respectively;
$\vert x\vert$ denotes the amplitude
of a complex scalar $x$; 
$\text{diag}(\cdot )$ is the diagonal operation;
$\mathcal{CN}\left(\mu, \sigma^2 \right)$ represents the complex univariate Gaussian distribution with the mean $\mu$ and the variance $\sigma^2$;
$\VM(\mu, \kappa)$ denotes the von Mises distribution with circular mean $\mu\in [0,2\pi]$ and centrality $\kappa$;
$\nc_{\chi_k^2}(\lambda, \sigma^2)$ and $\nc_{\chi_k}(\lambda)$ are the degree-$k$ non-central chi-squared distribution with  non-centrality parameter $\lambda$ and variance parameter $\sigma^2$, and the non-central chi distribution with non-centrality parameter $\lambda$, respectively; 
$\bm I_{L}$ denotes the $L\times L$ identity matrix;
$x_{\rm BB}(t)$ denotes the baseband representation of a passband signal $x(t)$;
$I_\nu(z)$ is the $\nu$-th order modified Bessel function of the first kind.

\section{System Model}  \label{System Model}
    \red{In this section, we will first specify the system model of the RIS-assisted \ac{MISO} system.
    Then, we will clarify the power allocation between CSI acquisition and data transmission in~\ref{Sec2-Subsec1}. 
    Finally, traditional approaches for the corresponding beamforming design and channel estimation will be introduced in Subsection~\ref{Beamforming design} and~\ref{Channel Estimation}, respectively.}

    Let us consider an RIS-aided MISO system, where an $N$-element RIS is employed for enhancing the transmission from an $M$-antenna \ac{BS} to a single-antenna user. 
    Assume furthermore that the phase-shift of each element of the RIS can be continuously and independently controlled with unit reflective gain~\cite{wu2019intelligent}. Then, the precoding matrix of the RIS can be represented by $\bm \Theta = \diag \left(\bm \theta\right)=\diag \left(\left[\theta_{1},\cdots ,\theta_{N}\right]\T\right)$,
    where $\theta_n (n\in \{N\})$ denotes the phase-shift of the $n$-th RIS element, satisfying $\lvert \theta_n\rvert=1$. Therefore, the signal received by the user can be written as 
        \begin{equation}
            \label{Signal model}
            y=\bm f\H \bm\Theta \bm G \bm w s+z,
        \end{equation}
        where $\bm f\in \mathbb C ^{N\times 1}$ and $\bm G \in \mathbb C^{N\times M}$ denote the channel spanning from the RIS to the user and the channel spanning from the BS to the RIS, respectively; $\bm w\in \mathbb C^{M\times 1}$ denotes the beamformer at the BS transmitter, with power constraint $\left\Vert \bm w\right\Vert_{2}^{2}\leq P_{\text{max}}$; $s$ denotes the normalized BS transmitted symbol satisfying $\mathbb{E}[ss^*]=1$; $z\sim \mathcal{CN}\left(0,\sigma_{z}^{2}\right)$ denotes the \ac{AWGN} imposed at the user's receiver. To focus on RIS beamforming, other possible links are neglected in this paper\footnote{In fact, our proposed CSI acquisition procedure automatically works when there exist direct BS-user links, which will be explained in Sec.~\ref{Sec4-Subsec4}.}.

\red{
\subsection{Power Allocation}\label{Sec2-Subsec1}

We assume a block-fading Rayleigh channel with $B$ symbols in each block, i.e., the i.i.d. Rayleigh random channel matrices $\bm G$ and $\bm f$ are updated for every consecutive $B$ symbols. Within these $B$ symbols, there are $N_p$ pilot symbols (i.e., the number of time slots for pilot signals) for channel estimation, and $N_d$ data symbols for data transmission, where $B=N_p+N_d$. The pilot symbols and the data symbols are subject to a total energy budget of $E_p$ and $E_d$, respectively.
Suppose the average transmit power of each block is $P$, then the energy values $E_p$ and $E_d$ should satisfy 
\begin{equation}
    P=\frac{E_p+E_d}{B}.
\end{equation}
Thus, a smaller number of pilot symbols $N_p$ will lead to a higher average pilot signal-to-noise ratio (SNR) $\gamma_p = E_p/(N_p \sigma_z^2)$ during CSI acquisition. This property ensures fair comparison among different channel estimation schemes, since equal energy $E_p$ is injected into the channel for the purpose of CSI acquisition.    

By defining the total pilot energy $E_p$ and the total data enery $E_d$, the actual transmit power values during the CSI acquisition phase and the data transmission phase are given by   
\begin{equation}
    \begin{aligned}
        P_{{\rm BS}, p} &= \frac{E_p}{N_p}P_{\rm max},\quad P_{{\rm BS}, d} = \frac{E_d}{N_d}P_{\rm max}, \\
        P_{{\rm u}, p} &= \frac{E_p}{N_p}P'_{\rm max},\quad P_{{\rm u}, d} = \frac{E_d}{N_d}P'_{\rm max},
    \end{aligned}
    \label{eqn:transmit_power}
\end{equation}
where $P_{\rm max}$ is the average BS transmit power budget, $P'_{\rm max}$ is the average user transmit power budget. 
}

\red{
\subsection{Beamforming Design} \label{Beamforming design}
    Based on the received signal~\eqref{Signal model}, we can formulate the \ac{SNR} maximization problem for designing the BS beamformer $\bm w$ and precoding matrix $\bm \Theta$, i.e.,
    \begin{subequations}
        \label{optimization}
        \begin{align}
        \label{objective}
            \max_{\bm \Theta}~~&\text{SNR}=\frac{1}{\sigma_{z}^{2}}
            \left\vert
            \bm f\H\bm \Theta\bm G\bm w \right\vert^{2},\\
        \label{constraint}
            ~~~~~\text{s.t.~~~}&\text{C}_{1}: \left\Vert \bm w\right\Vert_{2}^{2}\leq P_{\text{max}},\\
            &\text{C}_{2}: \left\vert\theta_{n}\right\vert=1,~\forall n.
        \end{align}
    \end{subequations}

    A near-optimal solution to this joint active-passive beamforming problem can be solved by the following alternating optimization (AO) loop~\cite{wu2019intelligent}:
    \begin{equation}
        \left\{\begin{aligned}
            & {\bm \theta}^{(t+1)} = \exp(-\ri \arg({\rm diag}({\bm f}^*) {\bm G} {\bm w}^{(t)})), \\
            & {\bm w}^{(t+1)} = \sqrt{P_{\rm max}}{\rm norm}(({{\rm diag}({\bm f}^*) {\bm G}})\H {\bm \theta}^{(t+1)*}),
        \end{aligned}\right. \label{eqn:iter_beamforming}
    \end{equation}
    where ${\rm norm}({\bm z}), {\bm z}\in\mathbb{C}^{M\times 1}$ denotes ${\bm z}/\|{\bm z}\|$, and $t$ is the iteration index. Note that this AO-based beamforming method depends on the knowledge of the full channel matrices $\bm G$ and $\bm f$, thus the active beamformer $\bm w$ and the passive beamformer $\bm \Theta$ have to be frequently re-optimized once the channel changes.  

    However, a more often assumption in real-world RIS-aided systems is that the locations of the \ac{BS} and the RIS are usually fixed, implying that the BS-RIS channel $\bm G$ enjoys a much longer channel coherence time compared with the RIS-user channel $\bm f$~\cite{Huchen}. Thus, in order to mitigate the computational cost, a suboptimal solution can be immediately obtained by designing $\bm w$ and $\bm\Theta$ in a one-shot manner, i.e., 
    \begin{equation}
        \begin{aligned}
            {\bm w} &= \sqrt{P_{{\rm BS}, d}}\underset{\tilde{\bm w}\in\mathbb{C}^M, \|\tilde{\bm w}\|=1}{\rm argmax}\,\|{\bm G}\tilde{\bm w}\|^2, \\
            \theta_{n} &= \exp\left(-\ri \arg\left(f_{n}^{*}\bm g_{n}\T \bm w\right)\right),~\forall n\in \{N\},
        \end{aligned}
        \label{traditional beamforming}
    \end{equation}
    where $\bm G = \left[\bm g_{1}, \cdots, \bm g_{N}\right]\T$. Note that this method is equivalent to optimizing the upper bound of the target function~\eqref{objective}:
    \begin{equation}
        {\rm SNR} = \frac{1}{\sigma_z^2}\left|\langle {\bm \Theta}{\bm G}{\bm w}, {\bm f}\rangle \right|^2 \leq \frac{1}{\sigma_z^2}\|{\bm G}{\bm w}\|^2\cdot \| {\bm f} \|^2.
    \end{equation} 
    }

\subsection{Channel Estimation}
\label{Channel Estimation}
    Observe from~\eqref{eqn:iter_beamforming} that, it suffices to know the cascaded channel ${\bm H} = {\rm diag}({\bm f}^{*})\bm G$ for beamforming design. 
    To accurately acquire this cascaded channel, the user sends pilot signal $x\in\mathbb{C}$ to the BS with $N_p$ different RIS configurations. These different RIS configurations are usually designed to be the first $N_p$ columns of the DFT matrix ${\bm F}_N$, denoted as $\bm F_{N,N_p}$. 
    Due to the channel reciprocity, the received signal at the BS with the $p$-th RIS configuration $\bm \Theta_{p}=\diag({\bm \theta}_p)$ can be written as~\cite{atapattu2020reconfigurable}
    \begin{equation}
    \label{CE received signal}
        {\bm y}_{{\rm BS}, p} =\bm H\T {\bm \theta}_p w' s'+\bm n,\quad p=1,2,\cdots,N_p,
    \end{equation}
    where $w'$ is the precoding scalar of the user satisfying $|w'|^2 \leq P_{\rm max}'$, $s'$ is the normalized user transmitted symbol with $\mathbb{E}[s'(s')^*]=1$, and $\bm n\sim \mathcal{CN}\left( \bm 0_{M}, \sigma_{z}^{2}\bm I_{M}\right)$ is the \ac{AWGN} at the BS receiver.
    Equivalently, \eqref{CE received signal} can be expressed in the matrix form
    \begin{equation}
    \label{eqn:LS_CE model}
        {\bm Y}_{\rm BS}=\bm H\T \bm F_{N,N_p}w' s' + {\bm N},
    \end{equation}
    where ${\bm Y}_{\rm BS}=\left[\bm y_{{\rm BS}, 1},\cdots,\bm y_{{\rm BS}, N_p}\right]$ and $\bm N = \left[ \bm n_{1},\cdots,\bm n_{N_p}\right]$.
    \red{Given the received signal $\bm Y_{\rm BS}$ and, without loss of generality, assuming $w's'=\sqrt{P'_{\rm max}}$, the channel estimation problem in RIS-assisted system can be solved by the MMSE estimator~\cite{kundu2021channel} as  
    \begin{equation}
        \begin{aligned}
            \hat{\bm H}\T &= \underset{\hat{\bm H}\T}{\rm argmin}\,\mathbb{E}\left[\|\hat{\bm H}\T - {\bm H}\T\|_{F}^2 | {\bm Y}_{\rm BS}\right]\\
            &= \mathbb{E}\left[ {\bm H}\T | {\bm Y}_{\rm BS} \right] 
        \end{aligned} 
        \label{eqn:MMSE_H}
    \end{equation}
    However, the evaluation of the MMSE estimator requires exact knowledge of the prior p.d.f. of the channel $p({\bm H})$, which is difficult to be written in an explicit form. This is because the entries of the cascaded channel matrix ${\bm H}_{ij}$ is the product of two Gaussian-distributed random variables ${\bm f}_i^{*}\sim{\mathcal{CN}}(0,\sigma_f^2)$ and ${\bm G}_{ij}\sim{\mathcal{CN}}(0,\sigma_g^2)$, which is generally not a Gaussian random variable.  }

    \red{An alternative way to circumvent this difficulty is to constrain the estimator $\hat{\bm H}$ in~\eqref{eqn:MMSE_H} to be a linear transform of ${\bm Y}_{\rm BS}$. This linear constraint produces a linearly approximated MMSE estimator, which is called the linear MMSE (LMMSE) estimator. By assuming $\sigma_h^2 = \sigma_f^2\sigma_g^2$, the LMMSE estimator can be obtained as }
    \red{
    \ifx\onecol\undefined
        \begin{equation}
            \begin{aligned}
            & (\widehat{\bm{H}\T})_{\rm LMMSE} \\
            & = \sqrt{P'_{\rm max}} {\bm Y}_{\rm BS} {\bm F}_{N, N_p}\H \left( P'_{\rm max}{\bm F}_{N, N_p} {\bm F}_{N, N_p}\H + \frac{\sigma_z^2}{\sigma_h^2}{\bm I}_{N}  \right)^{-1},
            \end{aligned}
            \label{eqn:LMMSE}
        \end{equation}
    \else
        \begin{equation}
            \widehat{\bm{H}\T}_{\rm LMMSE} = \sqrt{P'_{\rm max}} {\bm Y}_{\rm BS} {\bm F}_{N, N_p}\H \left( P'_{\rm max}{\bm F}_{N, N_p} {\bm F}_{N, N_p}\H + \frac{\sigma_z^2}{\sigma_h^2}{\bm I}_{N}  \right)^{-1},
            \label{eqn:LMMSE}
        \end{equation}
    \fi
    }
    \red{which provides a feasible solution to the channel estimation problem in the RIS-assisted MISO system. In this paper, we name it the linear minimum mean squared error for $\bm H$ (LMMSE-$\bm H$) channel estimation method. 

    Furthermore, by assuming known BS-RIS link $\bm G$, the channel estimation problem can be significantly simplified to the estimation problem of ${\bm f}^*$. By exchanging the position of ${\bm\theta}$ and ${\bm f}^*$ in~\eqref{CE received signal}, we obtain the equivalent uplink channel estimation model as 
    \begin{equation}
        {\bm y}_{{\rm BS}, p} = \sqrt{P'_{\rm max}} {\bm G}\T \diag({\bm \theta}_p) {\bm f}^* + {\bm n}, 
    \end{equation} 
    which immediately leads to the following MMSE\footnote{This LMMSE estimator is automatically a true MMSE estimator, since the model is linear, and the prior on $\bm f$ and the noise $\bm n$ is Gaussian. } estimator for $\bm f$:
    \begin{equation}
        \hat{\bm f}^* = \left( {\bm A}\H{\bm A} + \frac{\sigma_z^2}{\sigma_f^2}{\bm I}_N \right)^{-1} {\bm A}\H \tilde{\bm y}, \label{eqn:MMSE-f-estimator}
    \end{equation}
    where $\tilde{\bm y} = {\rm vec}({\bm Y}_{\rm BS})\in\mathbb{C}^{N_pM\times 1}$, and 
    \begin{equation}
        {\bm A} = \sqrt{P'_{\rm max}}\left[ {\bm G}\T\diag({\bm \theta}_1);\cdots;  {\bm G}\T\diag({\bm \theta}_{N_p})\right] \in \mathbb{C}^{N_pM\times N}.
    \end{equation}
    Since this channel estimation scheme assumes known $\bm G$ and estimates the unknown $\bm f$ by an MMSE estimator, we refer to it as the MMSE-$\bm f$ scheme. 

    It is worth noting that, both the LMMSE-$\bm G$ and the MMSE-$\bm f$ methods consume a linear number of pilot slots in $N$. 
    For the LMMSE-$\bm G$ scheme, 
    In order to determine the RIS-user channel $\bm{f}\in\mathbb{C}^{N\times 1}$, at least $N_p = \lceil N/M \rceil = {\Theta}(N)$ pilots are required, even with the assumption that the slow-varying~\cite{Huchen} BS-RIS channel $\bm G$ is known in advance. 
    In fact, without exploiting the inherent structure of $\bm f$, it seems impossible to obtain the $N$ channel coefficients with pilot overhead smaller than $\Theta(N)$. 
    However, in the following section, we will show that it is possible to realize dimension-independent CSI acquisition by exploiting a common physical phenomenon in electrodynamics. }
    
\section{Interference Random Field}
\label{Interference Random Field}
    In this section, the signal model for the \ac{IRF} will be first introduced in Subsection~\ref{Models for IRFs}. Then, in order to acquire the CSI, the simultaneous rotational signaling method will be proposed in Subsection~\ref{Simultaneous Rotational Signaling and Interference Detection}. 

\subsection{Signal Model for IRF}
\label{Models for IRFs}

    Physical intuition is vital in designing RIS-aided systems~\cite{najafi2020physics}, and interference is a fundamental physical phenomenon that appears in all kinds of waves. The most well-known example is the double-slit optical interference~\cite{louradour1993interference}, as is shown in Fig.~\ref{fig:scheme}~(a), where the interference fringes created during the wave superposition reveal the phase difference of the two optical paths. Similarly, the same interference phenomenon occurs when two RF electromagnetic waves meet together on the RIS, as is shown in Fig.~\ref{fig:scheme}~(b), and the \ac{IRF} created by this interference reveals the \ac{CSI}. 
    \ifx\onecol\undefined
    \begin{figure}[t]
        \centering
        \subfigure[Optical interference.]
        {\includegraphics[width=1\linewidth]{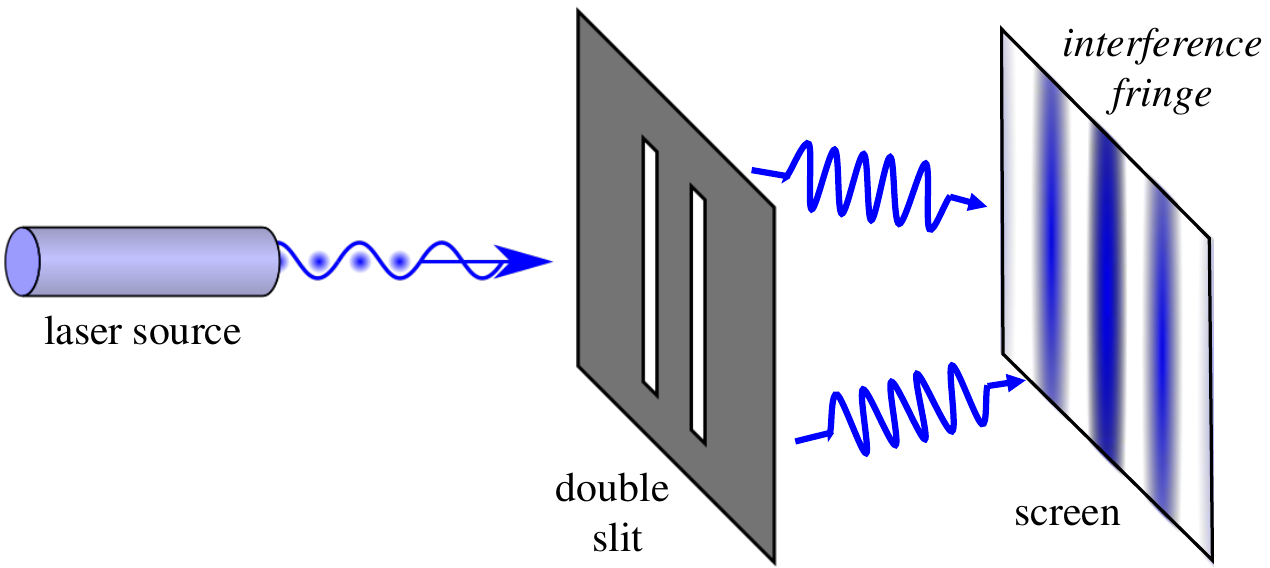}}
        \subfigure[Electromagnetic interference.]
        {\includegraphics[width=0.7\linewidth]{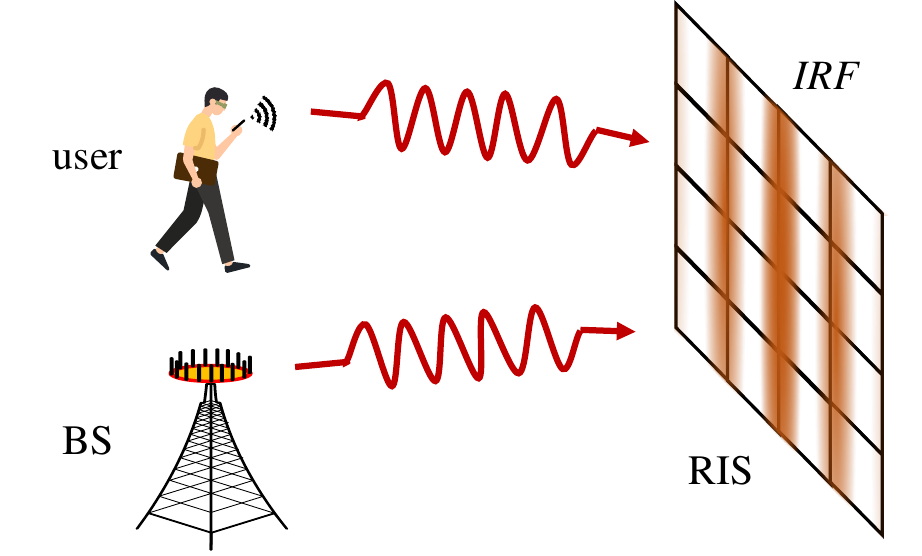}}
        \caption{Analogy between the optical interference and the \ac{IRF} phenomenon induced on the RIS.}
        \label{fig:scheme}
    \end{figure}
    \else 
        \begin{figure}[t]
            \centering
            \subfigure[Optical interference.]
            {\includegraphics[width=.58\textwidth]{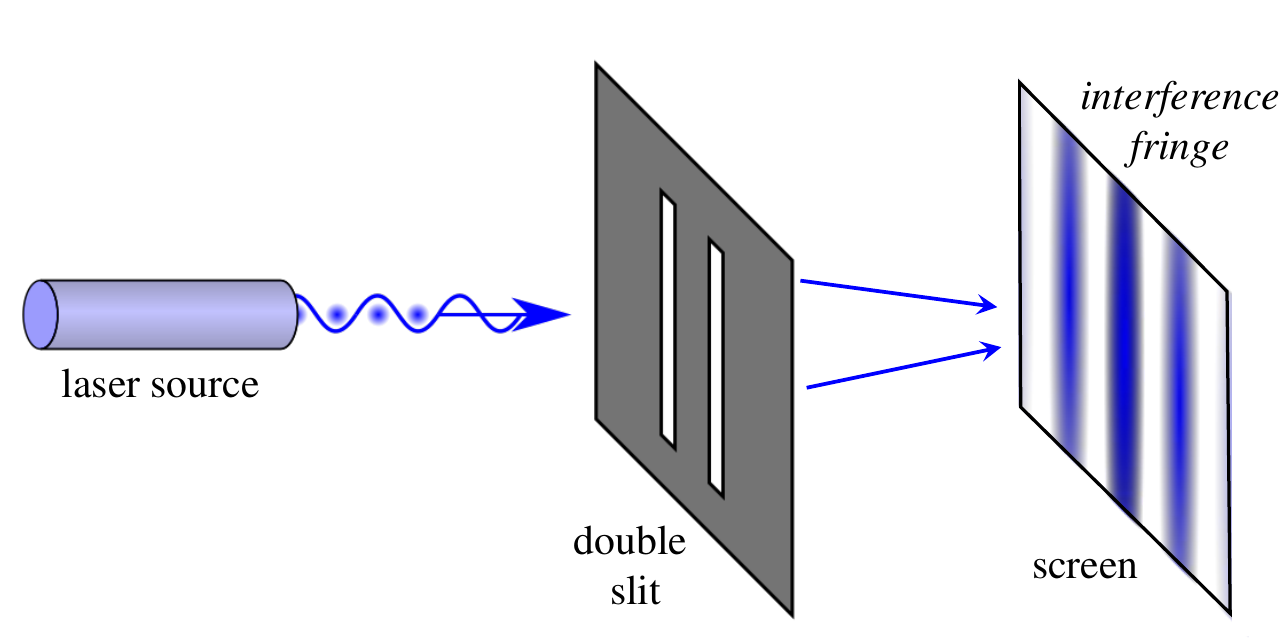}}
            \subfigure[Electromagnetic interference.]
            {\includegraphics[width=.41\textwidth]{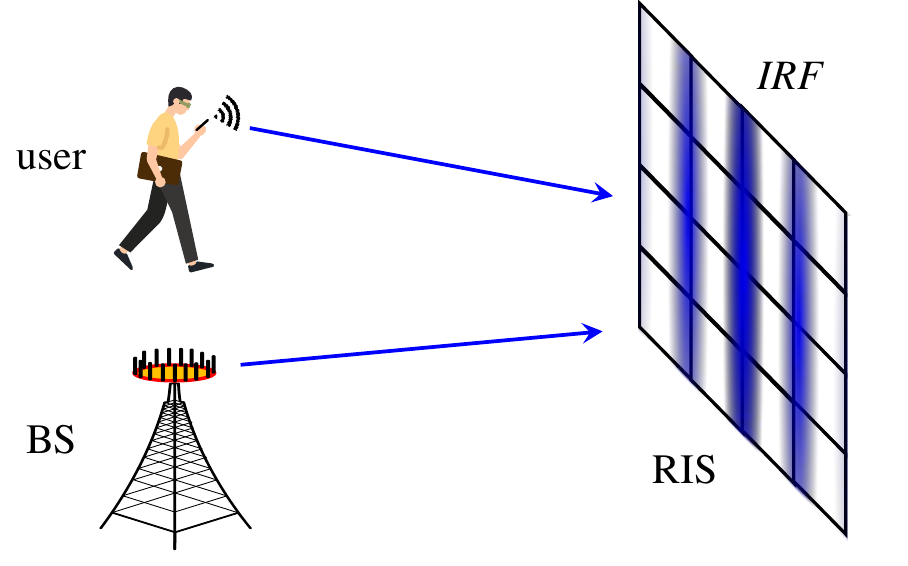}}
            \caption{Analogy between the optical interference and the \ac{IRF} phenomenon induced on the RIS.}
            \label{fig:scheme}
        \end{figure}
    \fi

    Suppose two signals impinge upon the RIS simultaneously, creating an interference field at the $n$-th RIS element. Denote the symbol transmitted from the \ac{BS} by $s$, and the symbol transmitted from the user by $s'$. In order to perform channel estimation by exploiting the \ac{IRF}, we need to probe the interference fringes, i.e., the power of the \ac{IRF}. To clarify the power problem associated with an electromagnetic signal, the relationship between physical signal power and baseband equivalent signal power is stressed in the following formulas. 
    If we represent both of the BS-RIS and user-RIS signals that appear on each RIS element by their baseband equivalent signals $E_{{\rm BB}}(t)$, 
    then the corresponding physical passband electric field induced on each RIS element is characterized by 
    \begin{equation}
        E(t) = \sqrt{2} \Re\left( E_{{\rm BB}}(t)e^{\ri\omega_c t} \right),
        \label{baseband-passband relationship}
    \end{equation}
    where $\omega_c = 2\pi f_c$ denotes the carrier frequency, and the coefficient $\sqrt{2}$ ensures that the passband signal power is equal to the baseband signal power, i.e., 
    $\lVert E_{{\rm BB}}\rVert_2^2 = \lVert E\rVert_2^2$. 
    Thus, taking the square of the baseband signal is equivalent to calculating the power of the physical electromagnetic signal. 
    Now we consider the IRF case, where a superposition of the BS-RIS signal and the user-RIS signal is considered. 
    Following the notations in Section~\ref{System Model} and due to the linearity of \eqref{baseband-passband relationship}, we obtain the noisy IRF signal by adding up the two impinging baseband signals
    \begin{equation}
        \begin{aligned}
            E_{{\rm BB, IRF}}(t) &= E_{{\rm BB, BS}}(t)+E_{{\rm BB,user}}(t)+v(t)\\
            &=\bm g_{n}\T\bm w s+f_{n}^{*} w' s' +v(t),
        \end{aligned}
    \end{equation}
    where $v(t)\sim \mathcal{CN}(0,\sigma_{v}^{2})$ is the electromagnetic noise signal in its baseband representation,
    $\bm w$ and $w'$ are the beamformer at the BS side and the user side satisfying $\Vert \bm w \Vert_{2}^{2}\leq P_{\text{max}}$ and $|w'|^{2}\leq P'_{\text{max}}$, respectively,
    $s$ is the symbol transmitted from the BS to RIS, and $s'=e^{\ri \psi(t)}$ is the time-varying transmitted symbol from the user to the RIS.  
    Since the interference fringes are not sensitive to a global phase change, we can safely assume that $s=1$, and then the relative phase between the user and the BS can be fully characterized by a time-varying phase function $\psi(t)$.  

    Furthermore, by defining $\alpha = \left\vert\bm g_{n}\T\bm w\right\vert$, $\beta = \left\vert f_{n}^{*} w' \right\vert$, and the phase difference between the \ac{BS}-RIS link\footnote{Since it is usually difficult to obtain the optimal BS beamformer $\bm w$, during the IRF-based CSI acquisition procedure, we fix $\bm w$ to a vector that guides most of the signal energy to the RIS aperture according to~\eqref{traditional beamforming}. As a result, $\bm g_n\T \bm w$ can be treated as the equivalent BS-link from the BS to the $n$-th RIS element.} and the RIS-user channel as $\varphi = \arg\left(f_{n}^{*}w'\right)-\arg\left(\bm g_{n}\T\bm w\right)$, then the power of the \ac{IRF} can be written as
    \ifx\onecol\undefined
        \begin{equation}
            \label{eqn:power of interference}
            \begin{aligned}
                P(t)&=A \left| E_{{\rm BB, IRF}} (t)\right |^{2}+\zeta\\
                &=\underbrace{A\left[\alpha^{2}+\beta^{2}+2\alpha\beta\cos\left(\psi(t)+\varphi\right)\right]}_{\text{IRF power signal}}\\
                &+\underbrace{2A\Re\left\{\left(\alpha+\beta e^{\ri\left(\psi(t)+\varphi\right)}\right)v'^{*}(t)\right\}+A\left|v'(t)\right|^{2}+\zeta}_{\text{Noise}},\\
            \end{aligned}
        \end{equation}
    \else 
        \begin{equation}
            \label{eqn:power of interference}
            \begin{aligned}
                P(t)&=A \left| E_{{\rm BB, IRF}} (t)\right |^{2}+\zeta\\
                &=\underbrace{A\left[\alpha^{2}+\beta^{2}+2\alpha\beta\cos\left(\psi(t)+\varphi\right)\right]}_{\text{IRF power signal}}+\underbrace{2A\Re\left\{\left(\alpha+\beta e^{\ri\left(\psi(t)+\varphi\right)}\right)v'^{*}(t)\right\}+A\left|v'(t)\right|^{2}+\zeta}_{\text{Noise}},\\
            \end{aligned}
        \end{equation}
    \fi
    where $\zeta$ is the noise introduced by digital signal processing after measuring the IRF power, $A$ is the amplification factor of the power sensor, and the equivalent electromagnetic noise $v'(t) = e^{\ri\cdot{\rm arg}({\bm g}_n\T \bm w)}v^*(t)$ follows the same distribution as $v(t)$ for any time $t$.
    The name {\it interference random field} (IRF) comes from the interferential nature of the electromagnetic field, the randomness of the unknown phase difference $\varphi$, and the unknown random noise realizations. 

    The phase difference $\varphi$ defined here will play an important role in our following channel estimation, since it carries enough CSI for beamforming, of which the reason will be justified in the next section. To estimate the phase difference $\varphi$, we need enough observations from the detected power signal $P(t)$. Recall that according to~\eqref{eqn:power of interference}, the interference power $P(t)$ is determined by $\alpha, \beta$ and the phase difference $\psi(t)+\varphi$. For simplicity and without loss of generality, we can assume $\psi(t)=\frac{2\pi}{T_{s}} t$, where $T_s$ is the symbol period. Furthermore, assume that $L$ observations $P[l]=P(t_{l})$ located at equally-spaced instants $t_{l}=\frac{l}{L}T_{s}$ for all $l\in \{L\}$
    are used for estimating the phase difference $\varphi$.

\subsection{Simultaneous Rotational Signaling and Interference Detection} \label{Simultaneous Rotational Signaling and Interference Detection}
    Generally, signal processing with only amplitude signals $P(t)$ are called {\it noncoherent} detection. The target of RIS beamforming is to strengthen the signal at the user, hence it requires {\it coherent} signal combining at the user antenna.  
    As a result, for {\it noncoherent} detection, it seems that the lack of phase information makes beamforming impossible. 
    Fortunately, the physical phenomenon {\it interference} makes it possible to convert phase difference into intensity difference, allowing {\it noncoherent} devices to perform {\it coherent} detection. 
    \ifx\onecol\undefined
        \begin{figure}[!t]
            \centering
            \includegraphics[width=1\linewidth]{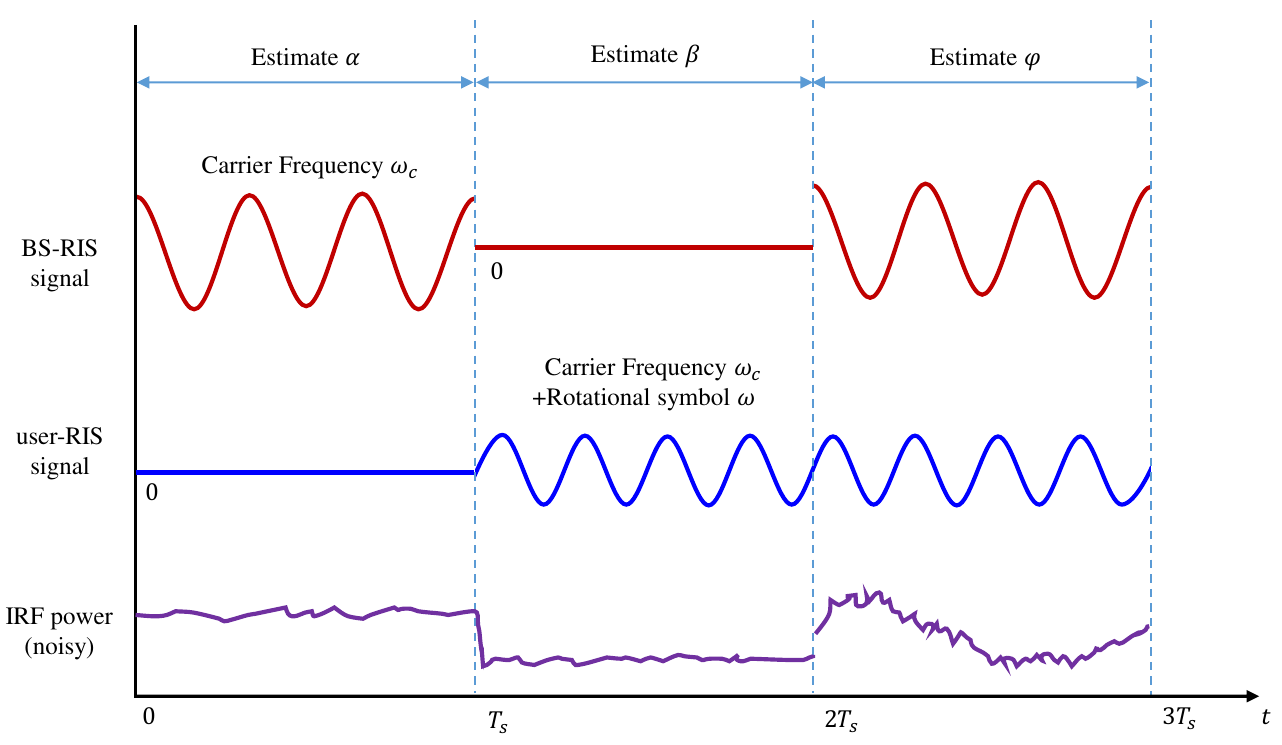}
            \caption{Simultaneous rotational signaling scheme. The IRF power is the instantaneous total power of the BS-RIS signal and the user-RIS signal. The composite power waveform that appears on RIS enables our algorithm to obtain the desired CSI. The parameters $\alpha$ and $\beta$ are estimated from the signals $P_{\alpha}(t), P_{\beta}(t)$ measured during the first two time slots, and $\varphi$ is estimated by IRF signal $P(t)$ that occurs during the third time slot.}
            \label{fig:protocol}
        \end{figure}
    \else 
        \begin{figure}[!t]
            \centering
            \includegraphics[width=.9\textwidth]{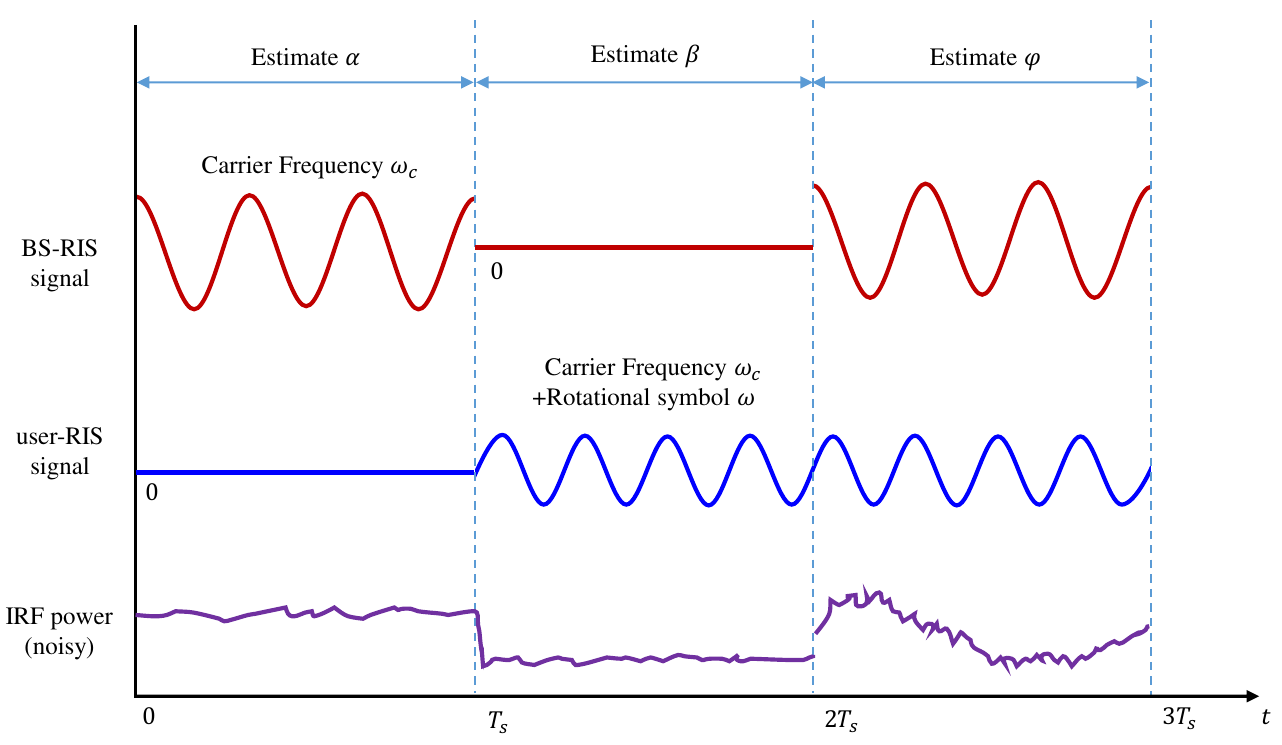}
            \caption{Simultaneous rotational signaling scheme. The IRF power is the instantaneous total power of the BS-RIS signal and the user-RIS signal. The composite power waveform that appears on RIS enables our algorithm to obtain the desired CSI. The parameters $\alpha$ and $\beta$ are estimated from the signals $P_{\alpha}(t), P_{\beta}(t)$ measured during the first two time slots, and $\varphi$ is estimated by IRF signal $P(t)$ that occurs during the third time slot.}
            \label{fig:protocol}
        \end{figure}
    \fi

    To clarify this idea, we take only one RIS element into consideration. 
    As is introduced in the previous subsection, if we allow the BS and the user to transmit electromagnetic waves simultaneously, then the interference phenomenon will occur on each RIS element. 
    However, equal carrier frequencies of the BS and the user create stable interference fringes~\cite{louradour1993interference} on the RIS, which do not carry information about the channel. 
    In contrast, a rotational symbol $s'=e^{\ri\psi(t)}$ at the user, which is equivalent to a slightly higher carrier frequency, enables the IRF power to vary over time. 
    Thus, the desired CSI can be drawn from the varying IRF power signal received by sensing RIS. 
    Fig.~\ref{fig:protocol} shows the simultaneous rotational signaling procedure and the IRF waveform. 

    \begin{figure}[!t]
        \centering
        \includegraphics[width=\linewidth]{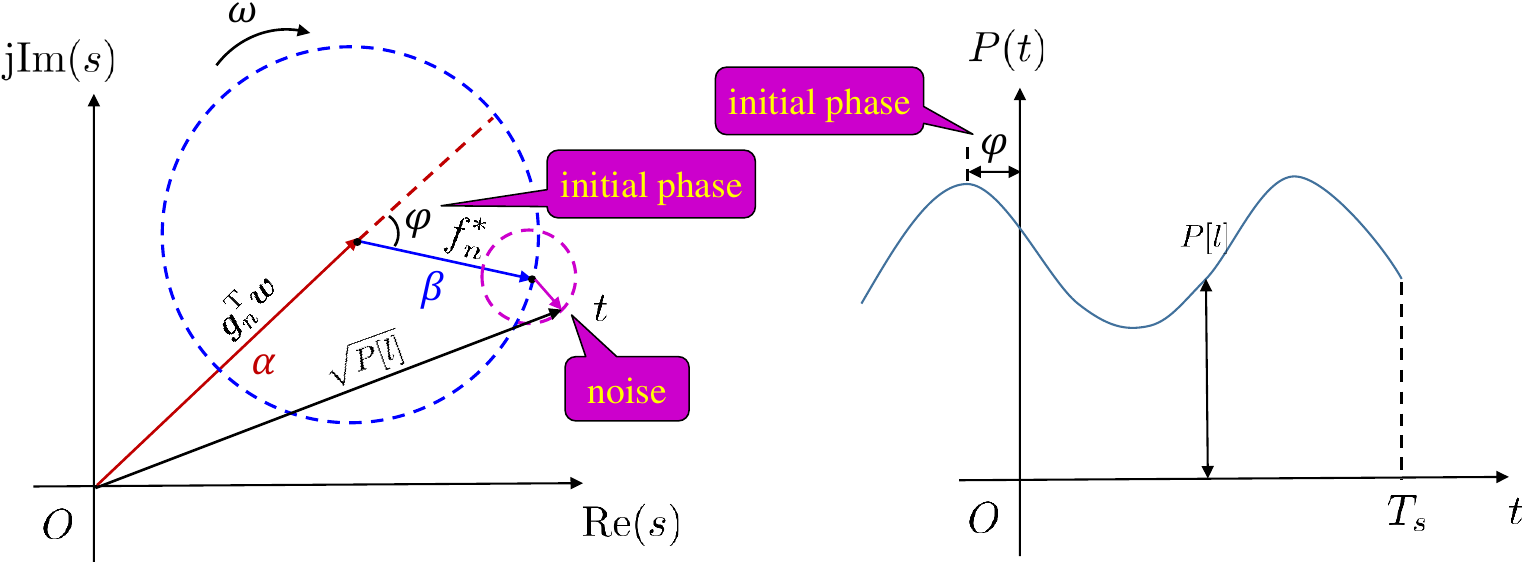}
        \caption{Phasor representation of the \ac{IRF} on each RIS element. Red vector represents the complex BS-RIS signal ${\bm g}_n\T{\bm w}$; Blue vector represents the complex user-RIS signal $f_n^*w'$; $\alpha$ and $\beta$ denote the amplitude of these signals respectively. In our signaling method, the BS transmits a fixed symbol $s=1$, while the user transmits a rotating symbol $s'=e^{\ri\psi(t)}$, causing the output of the power sensor $P(t)$ to vary in a waveform that is similar to a sine curve.}
        \label{fig:phasor}
    \end{figure}
    To further analyze the interference, we focus on the received IRF signal $P(t)$ of the $n$-th RIS element. 
    Note that the power of interference field measured at instant $t$, expressed by \eqref{eqn:power of interference}, exhibits a sinusoidal waveform in the time domain, as is shown in Fig.~\ref{fig:phasor}. 
    The initial phase ($t=0$) of this sinusoidal waveform is uniquely determined by $\varphi$.
    Thus, if we have access to the power received by the $n$-th element at successive instants $t_l$, it is possible to retrieve the phase difference $\varphi$. 
    However, according to~\eqref{traditional beamforming}, it is the phase sum of the RIS-user channel and the BS-RIS channel, i.e., $\arg({\bm g}_n\T{\bm w})+\arg(f_n^*w')$, that determines the optimal phase-shift of the $n$-th element. Given the phase difference $\varphi = \arg\left(f_{n}^{*}w'\right)-\arg\left(\bm g_{n}\T\bm w\right)$, we assume the phase $\arg({\bm g}_n\T{\bm w})$ to be known in our algorithms in order to acquire the phase sum.
    \red{The underlying reason for this assumption is that, in contrast to the fast time-varying RIS-user link, the BS-RIS link often exhibits a quasi-static property, thus two-timescale methods in~\cite{Huchen} can be applied, in which the quasi-static BS-RIS channel $\bm G$ is estimated in a longer timescale, while the frequently-varying RIS-user channel $\bm f$ is estimated in a shorter timescale.
    Since the BS-RIS link $\bm G$ is usually stable, it can be estimated only once for many following data frames. Thus, in our simulations for the IRF-based methods, the phase $\arg({\bm g}_n\T{\bm w})$ is assumed to be known.  }

\section{Sensing RIS-Based Channel Estimation}
\label{Sensing RIS-Based Channel Estimation}
\red{In this section, we will first briefly introduce the hardware architecture required for detecting the \ac{IRF} in Subsection~\ref{Hardware Architecture of Sensing RIS}. Then, the core concept of IRF channel estimation and beamforming procedures will be introduced in Subsection~\ref{IRF Channel Estimation and Beamforming}. After that, Subsection~\ref{Pilot Overhead} will include the analysis on pilot overhead and computational complexity of the proposed IRF-based CSI acquisition methods. Finally, the extension to the co-existence case of direct-reflective links are presented in Subsection~\ref{Sec4-Subsec4}. 
}

\subsection{Hardware Architecture of Sensing RIS} \label{Hardware Architecture of Sensing RIS}
\ifx\onecol\undefined 
    \begin{figure}[t]
        \centering 
        \includegraphics[width=0.7\linewidth]{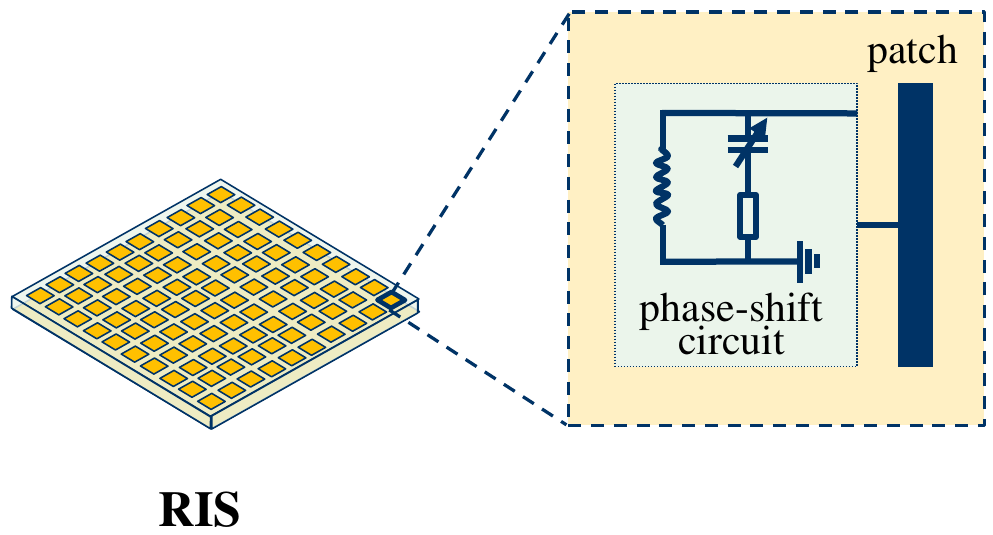}
        \centerline{\small (a) }
    \end{figure}
    \begin{figure}[t]
        \centering 
        \includegraphics[width=0.8\linewidth]{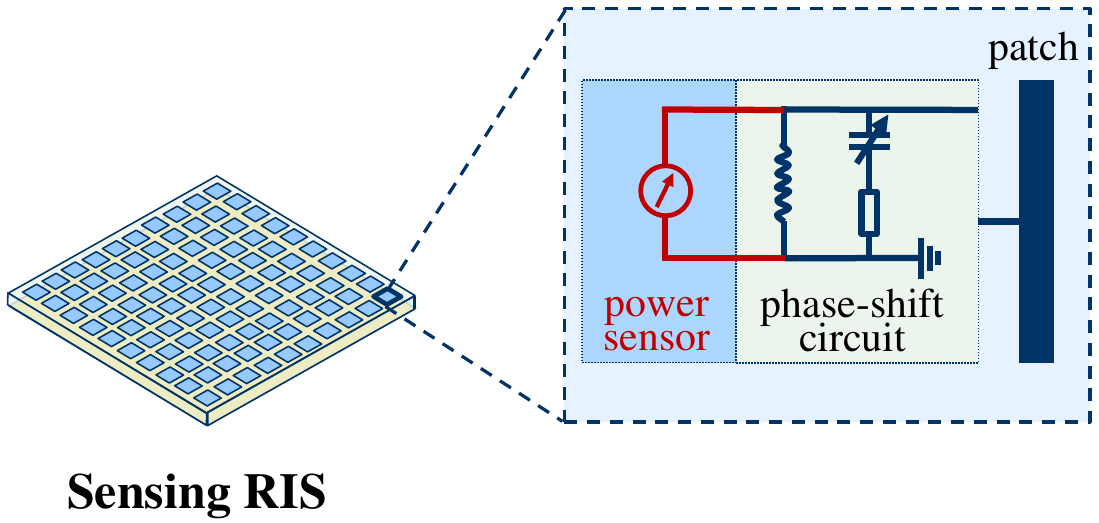}
        \centerline{\small (b) }
        \caption{Hardware structure comparison. (a) Traditional RIS. (b) Sensing RIS. }
        \label{fig:hardware}
    \end{figure}
\else 
\fi
    
To obtain the amplitude of the induced \ac{IRF}, inspired by the sensing metasurface~\cite{ma2019smart}, we propose a hardware architecture named sensing RIS.
In contrast to traditional RIS architecture shown in Fig.~\ref{fig:hardware}~(a), where each RIS element includes a phase-shift circuit and a patch antenna, each sensing RIS element additionally integrates a power sensor which is responsible for detecting the amplitude of the IRF~\cite{ma2020smartsensing}, as shown in Fig.~\ref{fig:hardware}~(b).
Since all of the RIS elements can sense and adjust the phase independently from one another, low-cost microcontroller units (MCUs) can be attached locally to each RIS element to allow parallel computation of the optimal phases. 
However, for extremely large-scale RIS systems, sparse sensing and controlling may be preferred to reduce cost and hardware complexity. 

\subsection{IRF Channel Estimation and Beamforming} \label{IRF Channel Estimation and Beamforming}
    In this subsection, we will thoroughly introduce the IRF-based algorithms for adjusting the phases of the RIS elements. The \ac{IRF} channel estimation and beamforming procedure can be divided into three steps:
    \begin{enumerate}
        \item Simultaneous rotational signaling and power sensing; 
        \item Phase estimation based on power data;
        \item Integrate phase information and other CSI to perform beamforming. 
    \end{enumerate} 

    In step 1), three power signals are recorded: $P_{\alpha}(t)$, $P_{\beta}(t)$, and $P(t)$ (see Fig.~\ref{fig:protocol} for details). The IRF appears during the third signal $P(t)$. 
    In order to create an IRF, simultaneous rotational signaling must be performed. 
    In fact, the signaling requirements can be realized by transmitting a symbol $s=1$ on the zeroth subcarrier at the BS, while transmitting a symbol $s'=1$ on the $k$-th subcarrier at the user. 
    As is mentioned above, if we denote the OFDM symbol period by $T_s$, then the equivalent IRF angular frequency $\omega$ is given by $\omega = 2\pi k /T_s$, which means the noiseless power signal exhibits exactly $k$ sinusoidal periods during an OFDM symbol. 
    Without loss of generality, in this paper, we always assume $k=1$. 
    \red{In fact, for multi-antenna MIMO systems, rotating pilot symbols at different frequencies $k$ can be simultaneously transmitted by different antennas to enable CSI acquisition.  
    Then, multi-antenna CSI acquisition can be fulfilled by performing Fourier analysis on the composite IRF power signal $P(t)$, and extracting the phase angles for each antenna at distinct frequencies. 
    Similarly, this pseudo-frequency division idea can be extended to multi-user MIMO systems, where different users can also be distinguished by different IRF frequencies. Fortunately, since different users are usually separated in the angular domain from the RIS's perspective, these users can be further identified by performing joint detection and estimation across all the RIS sensors. The joint CSI estimation is beyond the scope of this paper, and is left for our future work. }

    The second step 2) is called {\it phase estimation}. This step is designed to extract the phase difference $\varphi$ between the BS-RIS and RIS-user channel. 
    This is the key step of dimension-independent IRF channel estimation and beamforming, since this step can be done independently among all the RIS elements. 

    The third step 3) is to utilize the phase information provided by step 2) and calculate the near-optimal phase shifts for each RIS element in order to perform beamforming. The additional CSI refers to the phase information ${\rm arg}({\bm g}_n\T {\bm w})$. 

    Step 2) and step 3) are executed by processors, so we collect these two steps together into pseudo codes shown in {\bf Algorithm~\ref{alg:IRF-Beamforming}}:
    \begin{algorithm}[t] 
        \caption{Near-optimal RIS Beamforming by IRF} \label{alg:IRF-Beamforming}
        \setstretch{1.35}
        \begin{algorithmic}[1]
            \REQUIRE Number of RIS elements $N$, IRF power signals detected on each RIS element $P_{\alpha}(t), P_{\beta}(t)$ and $P(t)$.
            \ENSURE RIS phase-shift matrix ${\bm \Theta}$.
            \FOR{$n=1,2,\cdots,N$}
                \STATE Estimate $\alpha$ and $\beta$ from $P_{\alpha}(t)$ and $P_{\beta}(t)$.
                \STATE Estimate phase difference $\varphi_n$ from $P(t)$, $\alpha$ and $\beta$. 
                \STATE Estimate $\psi_{n} = {\rm arg}(\bm g_{n}\T\bm w)$ from known locations of BS and RIS
                \STATE $\theta_n \leftarrow \exp(-\ri(\varphi_n + 2\psi_n))$
            \ENDFOR
            \STATE ${\bm \Theta} \leftarrow \diag\left(\left[\theta_1, \theta_2, \cdots, \theta_N\right]\T\right)$
            \RETURN ${\bm \Theta}$
        \end{algorithmic}
    \end{algorithm}
    Note that in each iteration $n$ of {\bf Algorithm~\ref{alg:IRF-Beamforming}}, the data $P(t)$ is independent of other iterations. Thus, we can perform the calculations in parallel for each phase-shift $\theta_n$. For example, we can install an MCU for each of the elements on the RIS. Each MCU is only responsible for gathering the data from its own power sensor and adjusting the phase-shift of its own RIS element. Since all MCUs can work in parallel, the computational time is independent of the number of RIS elements, resulting in an $\mathcal{O}(1)$ time complexity.

\subsection{Pilot Overhead and Computational Complexity}\label{Pilot Overhead}
    In our proposed IRF-based CSI acquisition method, the pilot overhead is fixed to $\mathcal{O}(1)$, which is independent of the RIS dimension $N$. 
    The reason is that, no matter how many RIS elements are employed, the IRF appears on them simultaneously. 
    Thus, both the channel estimation and beamforming can be fulfilled within only three pilot symbols, as is depicted in Fig.~\ref{fig:protocol}. 
    Thus, the pilot overhead is independent of the RIS dimension $N$. 
    To the best of our knowledge, this dimension-independent property is unprecedented, if we cannot endure the cost of attaching a dedicated RF chain to every RIS element. Some hybrid solutions do exist, such as connecting all the RIS elements to several RF chains with analog combiners~\cite{alexandropoulos2020hardware,alexandropoulos2021hybrid} to enable explicit channel estimation at RIS. Unfortunately, the cost of analog combiners and RF chains are usually much higher than power sensors.
    
    \begin{table*}[th]
        \caption{Pilot Overhead Comparison of Different CSI Acquisition Methods}
        \label{tab:pilot overhead comp CE}
        \centering
        \begin{tabular}{|l|r|r|r|r}
            \hline 
            CSI acquisition method & Minimum pilot overhead per user & Type of RIS& Channel assumption \\ 
            \hline
            MVU\cite{jensen2020optimal}     & $N+1$    &RIS& General  \\
            \hline
            Multi-user\cite{wang2020channel}& $1+N/K+ \lceil \frac{(K-1)N}{M} \rceil/K$ &RIS& General  \\
            \hline
            CS\cite{wei2021channel}         & $\mathcal{O}(S\log N)$ &RIS& Sparse \\
            \hline 
            Two-timescale\cite{Huchen} & $\frac{2(N+1)}{\alpha K}+\lceil N/M\rceil+1$ &RIS& Quasi-static BS-RIS  \\
            \hline 
            Proposed IRF & $3$ & Sensing RIS  & Quasi-static BS-RIS\\ 
            \hline
        \end{tabular}
    \end{table*}

    \red{
    {\bf Table~\ref{tab:pilot overhead comp CE}} compares the pilot overhead of our proposed IRF method with other different CSI acquisition methods. 
    All the CSI acquisition methods that appear in {\bf Table~\ref{tab:pilot overhead comp CE}} assume a narrowband system, so the pilot overhead is equal to the number of pilot time slots.
    $K$ denotes the number of users, $S$ is the sparsity of the channel assumption, $M$ is the number of antennas at BS, and $\alpha$ is the ratio of the large-timescale channel  coherence time and the small-timescale channel coherence time~\cite{Huchen}. 
    Except for CS-based methods~\cite{wei2021channel}, all the channel estimation methods~\cite{jensen2020optimal,wang2020channel,Huchen} require the pilot overhead to be linearly dependent on $N$, while our IRF method needs a constant number of exactly 3 pilot slots per user, regardless of the number of RIS elements. 
    }

    The computational complexity of processing the obtained power signals depends on the hardware implementation. 
    The $\mathcal{O}(1)$ time complexity only holds when one MCU is installed for each RIS element. 
    If all the RIS elements are collectively controlled by one processor, the computational time would be $\mathcal{O}(N)$, but the pilot overhead still remains $\mathcal{O}(1)$. 
    These conclusions of complexity can be easily extended to multi-RIS schemes, where multiple RISs are employed to serve a single user at the same time. 
    The independent nature of IRF methods allows the RISs to work without the need to exchange data with the BS or other RISs. 
    This property makes it much easier to integrate a new RIS into an existing communication system, which greatly enhances the extendibility of the system.


\red{
\subsection{Extension to Systems with Direct BS-User Link} \label{Sec4-Subsec4}
Since the direct link does not affect the creation of the IRF, the RIS phase-shift matrix $\bm \Theta$ can be obtained as if there were no such a direct link, up to an undetermined global phase $e^{\ri \phi}$. Note that this IRF-based procedure for obtaining ${\bf\Theta}$ consumes only 3 time slots. After this procedure, two additional uplink training symbols can be transmitted from the user. The uplink channel model is 
\begin{equation}
    \begin{aligned}
        {\bm y}_{\rm BS} &= ({\bm h}_{\rm d} + e^{\ri \phi}{\bm G}\T {\bm \Theta} {\bm f}^*) s' + {\bm n} \\
        &= ({\bm h}_{\rm d} + e^{\ri \phi}{\bm h}_{\rm RIS})s' + {\bm n},
    \end{aligned}  
\end{equation}
where ${\bm h}_{\rm d}\in\mathbb{C}^{M\times 1}$ is the direct BS-user link. 
In the first symbol period $\phi$ is set to 0, and in the second $\phi=\pi$. Thus, the direct BS-user link ${\bm h}_{\rm d}$ and the reflective link ${\bm h}_{\rm RIS}$ can be recovered by 
\begin{equation}
    \begin{aligned}
        \hat{\bm h}_{\rm d} &= \frac{{\bm y}_{{\rm BS}, 1} + {\bm y}_{{\rm BS}, 2}}{2},\\
        \hat{\bm h}_{\rm RIS} &= \frac{{\bm y}_{{\rm BS}, 1} - {\bm y}_{{\rm BS}, 2}}{2}.
    \end{aligned}
\end{equation}
After obtaining the estimators of these two links, the global phase $\phi$ should be tuned to maximize the total channel energy $\| {\bm h}_{\rm d} + e^{\ri \phi}{\bm h}_{\rm RIS} \|^2$, and this is done by setting 
\begin{equation}
    \hat{\phi} = \arg(\hat{\bm h}_{\rm RIS}\H \hat{\bm h}_{\rm d}).
\end{equation}
Though the final solution $\exp(\ri \hat{\phi}) {\bm \Theta}$ for the RIS phase-shift matrix is suboptimal in general, it only consumes two additional time slots, resulting in 5 time slots in total for configuring $N$ reflective elements of the RIS. This pilot overhead is still dimension-independent. 
}

\section{Phase Estimation Algorithms}
In this section, three phase estimation algorithms will be proposed in Subsection~\ref{DFT method}, \ref{ML method}, and~\ref{von Mises-EM method} respectively, which constitute the core of the \ac{IRF} channel estimation and beamforming algorithm.

\subsection{DFT method}  \label{DFT method}
    The key challenge of the IRF channel estimation and beamforming is the phase estimation step, i.e., how to obtain the phase difference $\varphi$. Since the interferential power $P(t)$ exhibits a sinusoidal waveform, Fourier transforms can be applied to extract its phase. Apply $L$-point \ac{DFT} to the discrete-time  observed sensor detection signals $P[0],\cdots ,P[L-1]$, and we have
    \begin{equation}
        \label{DFT}
        p[l']=\sum\nolimits_{l=0}^{L-1}P[l]e^{-\ri\frac{2\pi}{L}ll'},\quad \forall l'\in \{L\}.
    \end{equation}
    Specifically, we have the complex amplitude of the first harmonic $p[1]$ as 
    \ifx\onecol\undefined
        \begin{equation} \label{DFT l=1}
            \begin{aligned}
                p[1]&=\sum\nolimits_{l=0}^{L-1}A\left[\alpha^{2}+\beta^{2}+2\alpha\beta\cos\left(\frac{2\pi}{L}l+\varphi\right)\right]e^{-\ri\frac{2\pi}{L}l}\\
                &=LA\alpha\beta  e^{\ri\varphi}.
            \end{aligned}
        \end{equation}
    \else 
        \begin{equation}
            \label{DFT l=1}
            p[1]=\sum\nolimits_{l=0}^{L-1}A\left[\alpha^{2}+\beta^{2}+2\alpha\beta\cos\left(\frac{2\pi}{L}l+\varphi\right)\right]e^{-{\ri}\frac{2\pi}{L}l}=LA\alpha\beta  e^{{\ri}\varphi}.
        \end{equation}
    \fi
    Then, the phase $\varphi$ can be estimated as
    \begin{equation}
        \label{LS estimate result}
        \hat{\varphi}=\arg\left(\frac{p[1]}{LA\alpha\beta}\right) = \arg\left(p[1]\right).
    \end{equation}
    Note that this DFT method simply ignores the non-Gaussian noise. For Gaussian noise, the DFT method is optimal. However, in fact, the noise in \eqref{eqn:power of interference} contains a squared term of Gaussian noise, resulting in the noise being non-Gaussian. Thus, we further conceive an ML method, as described in the following Subsection~\ref{ML method}. 

\subsection{Newton-ML method}  \label{ML method}
    Suppose the noise field $v'(t)=v'_R(t) + {\ri }v'_I(t)\sim \mathcal{CN}(0, \sigma_v^2)$, and the noise of the power sensor is $\zeta \sim \mathcal{N}(0, \sigma_{\zeta}^2)$. 
    Without loss of generality, we can assume that the sensor noise power $\sigma_{\zeta}^2$ is much weaker than the electromagnetic noise field $v'(t)$.
    Thus, we assume $\sigma_{\zeta}^2=0$ in the following discussion.  
    As a result, the distribution of $P(t) = A\left|E_{{\rm BB}, {\rm IRF}}(t)\right|^2$ is a non-central chi-squared distribution $\nc_{\chi_2^2}(A(\mu_{R}^2+\mu_{I}^2),  A\sigma_v^2/2)$ with degrees of freedom $k=2$, and mean values $\mu_{R}, \mu_{I}$ given by
    \begin{equation}
        \mu_{R} = \alpha + \beta \cos(\psi(t)+\varphi),\quad  \mu_{I}  = \beta \sin(\psi(t)+\varphi).
        \label{chi2 distribution mean values}
    \end{equation}
    Thus, according to \eqref{eqn:power of interference}, the output signal of the power sensor is given by 
    \begin{equation}
        P(t)  = A\left((v'_{R} + \mu_{R})^2 + (v'_{I} + \mu_{I})^2 \right)
        \label{eqn:sensor power}
    \end{equation}
    Let us define the noncentral parameter $\lambda(t)$ as
    \begin{equation}
        \lambda(t)  = A(\mu_{R}^2 + \mu_{I}^2) = A\left[\alpha^{2}+\beta^{2}+2\alpha\beta\cos\left(\psi(t)+\varphi\right)\right],
    \end{equation}
    then, according to the definition of the $\nc_{\chi_2^2}$, the p.d.f. of $P(t)$ is given by the zeroth-order modified Bessel function of the first kind 
    \begin{equation}
        f_{P}(x) = \frac{1}{A\sigma_{v}^2} \exp\left(-\frac{x+\lambda(t)}{A\sigma_v^2}\right)I_{0}\left(\frac{\sqrt{\lambda(t) x}}{A\sigma_v^2/2}\right),\quad x \geq 0.
        \label{ML single observation}
    \end{equation}
    Then, the log likelihood function of $\varphi$ based on the observations $P[l]$ can be represented by
    \ifx\onecol\undefined
        \begin{equation}
            \begin{aligned}
                & \mathcal{L}(P[0],\cdots,P[L-1] | \varphi) \\
                & = \sum_{l=0}^{L-1}\left[-\frac{P[l] + \lambda_l}{A\sigma_v^2} + \log I_0\left(\frac{\sqrt{P[l] \lambda_l}}{A\sigma_v^2/2}\right)\right] - L\log(A\sigma_v^2),
            \end{aligned}
            \label{ML likelihood}
        \end{equation}
    \else 
        \begin{equation}
            \mathcal{L}(P[0],\cdots,P[L-1] | \varphi) = \sum_{l=0}^{L-1}\left[-\frac{P[l] + \lambda_l}{A\sigma_v^2} + \log I_0\left(\frac{\sqrt{P[l] \lambda_l}}{A\sigma_v^2/2}\right)\right] - L\log(A\sigma_v^2),
            \label{ML likelihood}
        \end{equation}
    \fi
    where $\lambda_l := \lambda(t_l)$, and the derivative of \eqref{ML likelihood} is 
    \ifx\onecol\undefined
        \begin{equation}
            \begin{aligned}
                & \frac{\partial \mathcal{L}(P[0],\cdots,P[L-1] | \varphi)}{\partial \varphi} \\
                &= \frac{2\alpha\beta}{\sigma_v^2}\sum_{l=0}^{L-1}\sin(\psi(t_l)+\varphi) \left[1 - R\left( \frac{\sqrt{P[l]\lambda_l}}{A\sigma_v^2/2} \right) \frac{\sqrt{P[l]}}{\sqrt{\lambda_l}}\right],
            \end{aligned}
            \label{eqn:First Derivative Likelihood}
        \end{equation}
    \else 
        \begin{equation}
            \frac{\partial \mathcal{L}(P[0],\cdots,P[L-1] | \varphi)}{\partial \varphi} = \frac{2\alpha\beta}{\sigma_v^2}\sum_{l=0}^{L-1}\sin(\psi(t_l)+\varphi) \left[1 - R\left( \frac{\sqrt{P[l]\lambda_l}}{A\sigma_v^2/2} \right) \frac{\sqrt{P[l]}}{\sqrt{\lambda_l}}\right],
            \label{eqn:First Derivative Likelihood}
        \end{equation}
    \fi
    where the function $R(z)$ is defined as $R(z) = I_1(z)/I_0(z)$. Since the derivative of the function $R(z)$ satisfies the property \cite{silverman1972special}
    \begin{equation}
        R'(z)=1-R^2(z)-\frac{1}{z}R(z),
        \label{eqn:R function derivative property}
    \end{equation}
    the second derivative of the likelihood function $\mathcal{L}$ can be expressed as
    \ifx\onecol\undefined
        \begin{equation}
            \begin{aligned}
            & \frac{\partial^2 \mathcal{L}(P[0],\cdots,P[L-1] | \varphi)}{\partial \varphi^2} \\
            & =  \frac{2\alpha\beta}{\sigma_v^2} \sum_{l=0}^{L-1}{\cos(\psi(t_l)+\varphi)}\left[1 - R\left(z_l\right) \frac{\sqrt{P[l]}}{\sqrt{\lambda_l}}\right] \\
            & +\frac{4\alpha^2\beta^2}{\sigma_v^4}\sum_{l=0}^{L-1}{\sin^2(\psi(t_l)+\varphi) \left(1-R^2(z_l) -\frac{2}{z_l}R(z_l)\right)\frac{P[l]}{\lambda_l} },\\
            \end{aligned}
            \label{Second Derivative Likelihood}
        \end{equation}
    \else 
        \begin{equation}
            \begin{aligned}
            \frac{\partial^2 \mathcal{L}(P[0],\cdots,P[L-1] | \varphi)}{\partial \varphi^2}  = &  \frac{2\alpha\beta}{\sigma_v^2} \sum_{l=0}^{L-1}{\cos(\psi(t_l)+\varphi)}\left[1 - R\left(z_l\right) \frac{\sqrt{P[l]}}{\sqrt{\lambda_l}}\right] \\
            & +\frac{4\alpha^2\beta^2}{\sigma_v^4}\sum_{l=0}^{L-1}{\sin^2(\psi(t_l)+\varphi) \left(1-R^2(z_l) -\frac{2}{z_l}R(z_l)\right)\frac{P[l]}{\lambda_l} },\\
            \end{aligned}
            \label{Second Derivative Likelihood}
        \end{equation}
    \fi
    where $z_l = \sqrt{P[l]\lambda(t_l)}/(A\sigma_v^2/2)$.
    Then, we can perform the Newton iteration to obtain $\hat{\varphi}$, by iteratively using the updating formula
    \begin{equation}
        \hat{\varphi}^{(k+1)} = \hat{\varphi}^{(k)} - \frac{\mathcal{L}'(\hat{\varphi}^{(k)})}{\mathcal{L}''(\hat{\varphi}^{(k)})},
        \label{eqn:Newton-ML renewal formula}
    \end{equation}
    where $\mathcal{L}'(\hat{\varphi}^{(k)})$ and $\mathcal{L}''(\hat{\varphi}^{(k)})$ are given by \eqref{eqn:First Derivative Likelihood} and \eqref{Second Derivative Likelihood} respectively. Note that during the calculation of \eqref{eqn:Newton-ML renewal formula}, $A, \sigma_v, \alpha, \beta$ and the received signal $P[l]$ are all assumed to be known. Thus, in fact,  \eqref{eqn:First Derivative Likelihood} and \eqref{Second Derivative Likelihood} are functions of a single variable $\varphi$. 

\subsection{von Mises-EM method}    \label{von Mises-EM method}
    The Newton-ML algorithm, if convergent, is asymptotically optimal \cite{casella2021statistical}. 
    However, the computation of the Newton-ML estimator is quite complicated due to the intensive calculation of modified Bessel functions. 
    Now we introduce an iterative method for estimating $\varphi$ without any computation of such special functions.
    Our method is based on the von Mises distributions \cite{gatto2007generalized}.

    The von Mises-EM algorithm is based on the Bayesian inference of von Mises distributions \cite{mardia1976bayesian}. The von Mises distribution $\VM(\mu, \kappa)$ is a two-parameter distribution on $[0, 2\pi]$, with the probability density function given by 
    \begin{equation}
        p(\theta|\mu, \kappa) = \frac{\exp(\kappa \cos(\theta - \mu))}{2\pi I_0(\kappa)}, \quad 0\leq \theta \leq 2\pi,
    \end{equation}
    where $\mu \in [0,2\pi]$ and $\kappa >0$ being the cyclic location parameter and the concentration parameter. 
    Note that the von Mises distribution is a distribution on a circle, thus it acts as a perfect prior distribution of a phase estimation problem. 
    More fortunately, the von Mises distribution is also closely related to the complex Gaussian distribution, thus implying the possibility of designing an iterative EM algorithm \cite{casella2021statistical} based on the interactions between the von Mises distribution and the complex Gaussian noise distribution. The following two lemmas: {\bf Lemma \ref{lemma_1}} and {\bf Lemma \ref{lemma_2}} reveal these interactions. 
    \begin{lemma}[Bayesian estimation of $\VM$ distribution]\label{lemma_1} \mbox{}\par
        Let $\theta \sim \VM(\mu, \kappa)$, and $z|\theta \sim \CN(e^{{\ri}\theta}, \sigma^2)$. Then the posterior distribution $\theta | z$ is also a von Mises distribution $\VM(\mu', \kappa')$ with parameters $\mu'$ and $\kappa'$ satisfying $\kappa' e^{{\ri}\mu'} = \kappa e^{{\ri}\mu} + 2z/\sigma^2$.
    \end{lemma}
        \begin{IEEEproof}
        See {\bf Appendix \ref{Proof of Lemma 1}}. 
    \end{IEEEproof}
    
    In this paper, we also use $\VM(\kappa e^{{\ri}\mu})$ to denote the von Mises distribution $\VM(\mu, \kappa)$. 
    This representation provides convenience for the calculation of the posterior distribution of the von Mises distribution in Bayesian inference.

    \begin{lemma}[Circular $\CN$ posterior is $\VM$]\label{lemma_2} \mbox{}\par
        Suppose $z \sim \CN(z_0, \sigma^2)$, where $z_0 \in \mathbbm{C}$, and a positive radius $r>0$. Then the posterior distribution of angle $\theta= {\rm arg} (z)$, constrained on a circle $|z|=r$ obeys the von Mises distribution
        \begin{equation}
            p(\theta |\, |z|=r) \sim \VM\left({\rm arg}(z_0), \frac{r|z_0|}{\sigma^2/2}\right).
        \end{equation}
    \end{lemma}
    \begin{IEEEproof}
        Replacing $z$ by $r e^{\ri \theta}$ give rise to the conclusion immediately. 
    \end{IEEEproof}
    
    Combining the results of {\bf Lemma \ref{lemma_1}} and {\bf Lemma \ref{lemma_2}}, we can then construct the EM algorithm for estimating $\varphi$. Since the output of the power sensors $P[l]$ does not contain phase information, we can treat the phases as latent variables. Let $s_l = \sqrt{P[l]/A}$ be the noisy estimation for $|\alpha + \beta e^{{\ri} (\varphi + \psi_l)} + v_l|$, and $\theta_l$ be the latent variables $\arg (\alpha + \beta e^{{\ri} (\varphi + \psi_l)} + v_l)$ that are not observable. 
    Since the noise $v_l \sim {\rm i.i.d.}\;\CN(0, \sigma_v^2)$, then from {\bf Lemma \ref{lemma_2}}, $\theta_l | s_l, \varphi \sim \VM(\arg(\alpha + \beta e^{{\ri} (\varphi + \psi_l)}), s_l |\alpha + \beta e^{{\ri} (\varphi + \psi_l)}|/(\sigma_v^2/2))$. Thus, we can infer the latent variables by ML estimation 
    \begin{equation}
        \hat{\theta}_{l, {\rm ML}} | s_l, \varphi = \arg(\alpha + \beta e^{{\ri} (\varphi + \psi_l)}).
        \label{eqn:E-step}
    \end{equation}
    After inferring the latent variables $\hat{\theta}_{l, {\rm ML}}$, we can update the estimation of $\varphi$ using Bayesian rule in {\bf Lemma \ref{lemma_1}}
    \begin{equation}
        \varphi | s_l, \theta_l \sim \VM\left( \kappa e^{{\ri} \mu} + \frac{\beta}{\sigma_v^2/2}\sum_{l=0}^{L-1}{\left(s_l e^{{\ri}\theta_l}-\alpha\right)e^{-{\ri} \psi_l}}\right),
        \label{eqn:M-step}
    \end{equation}
    where the coefficient $\beta/(\sigma_v^2/2)$ comes from scaling the phasor in Fig.~\ref{fig:phasor} by a factor $\beta^{-1}$ and applying {\bf Lemma \ref{lemma_1}}.
    Performing E-step with \eqref{eqn:E-step} and M-step with \eqref{eqn:M-step} alternately, then the estimation precision for $\varphi$ can be iteratively improved. Note that although the modified Bessel functions appear in the density function of von Mises distribution, the bother is avoided in the von Mises-EM algorithm. The pseudo code of von Mises-EM algorithm is collected in {\bf Algorithm \ref{alg:vM-EM}}.

    \begin{algorithm}[t] 
        \caption{von Mises-EM phase estimation (vM-EM algorithm)} \label{alg:vM-EM}
        \setstretch{1.35}
        \begin{algorithmic}[1]
            \REQUIRE Incident wave intensity $\alpha$, $\beta$; sensor data $P[l]$; amplification factor $A$ and noise variance $\sigma_v^2$; predefined phase shifts $\psi_l=\omega t_l$.
            \ENSURE $\hat{\varphi}$
            \STATE $s_l \leftarrow \sqrt{P[l]/A}, \quad \forall l\in \{L\}$
            \STATE $\hat{\varphi} \leftarrow \arg\{{\rm FFT}(P)[1]\}$
            \STATE $\kappa \leftarrow 1$
            \WHILE {$\hat{\varphi}$ not convergent}
                \STATE $\mu_l \leftarrow \alpha + \beta e^{{\ri} (\hat{\varphi}+\psi_l)}, \quad \forall l\in \{L\}$
                \STATE $w_l \leftarrow s_l e^{{\ri} \arg(\mu_l)} - \alpha, \quad \forall l\in \{L\}$
                \STATE $z_\varphi \leftarrow \kappa e^{{\ri} \hat{\varphi}} + \beta \left( \sum_{l=0}^{L-1}{w_l e^{-{\ri} \psi_l}}\right) / (\sigma_v^2/2)$
                \STATE $\hat{\varphi} \leftarrow \arg(z_\varphi)$
            \ENDWHILE
            \RETURN $\hat{\varphi}$
        \end{algorithmic}
    \end{algorithm}
    
\section{Performance Analysis}
\label{Performance Analysis}
    In this section, we first provide the achievability proof the proposed vM-EM phase estimation algorithm. Then, we derive various CRLB expressions for the phase estimation problems. 

\subsection{Asymptotic Achievability Bound of the vM-EM algorithm}
\red{\begin{theorem}[Fixed-point perturbation bound] \label{thm:asymp_perf_vM-EM}
    Assume $\alpha>\beta>0$ are fixed. Then, there exists a sufficiently large integer $L_0$, such that for any $L\geq L_0$, if the proposed vM-EM algorithm converges, the returned estimator $\hat{\varphi}$ achieves an MSE performance of $\mathbb{E}[(\hat{\varphi}-\varphi)^2] = \mathcal{O}(\bar{\gamma}^{-1})$, where $\varphi$ is the true parameter, and $\bar{\gamma}$ is the interferential SNR defined as $\bar{\gamma} = (\alpha^2+\beta^2)/\sigma_v^2$.    
\end{theorem}

Before proving this theorem, we first introduce the following {\bf Lemma~\ref{lemma:wirtinger_der_vM-EM}} that characterizes how fast the vM-EM estimator $\hat{\varphi}$ in {\bf Algorithm~\ref{alg:vM-EM}} varies with the random noise vector $\bm v$.

\begin{lemma}[The speed of the fixed-point] \label{lemma:wirtinger_der_vM-EM}
    Suppose that for the input data sequence ${\bm s} = \sqrt{{\bm P}/A} = (s_0, \cdots, s_{L-1})\T$, the vM-EM algorithm converges to the estimator $\hat{\varphi}$. Then, the estimator $\hat{\varphi} = \hat{\varphi}({\bm v})$ can be viewed as a function of the noise $\bm v$, and the Wirtinger derivative~\cite{remmert1991theory} of $\hat{\varphi}$ w.r.t. $\bm v$ satisfies 
    \begin{equation}
        \|\nabla_{\bm v}\hat{\varphi}\|^2 = \frac{1}{|\langle {\bm s}, {\bm x}\rangle|^2}\sum_{\ell=0}^{L-1}  \frac{\sin^2(\theta_\ell)}{|\mu_\ell|^2}, 
    \end{equation} 
    where ${\bm x} = (x_0, \cdots, x_{L-1})\T$ is defined as 
    \begin{equation}
        x_{\ell} := \frac{(\alpha+\beta \cos\theta_\ell)(\beta+\alpha\cos\theta_\ell)}{|\mu_\ell|^3},~~\ell\in\{L\}, 
    \end{equation}
    and 
    \begin{equation}
        \begin{aligned}
            \mu_\ell & := \alpha+\beta {\rm exp}(\ri \theta_\ell),~~\ell\in\{L\}, \\
            \theta_\ell & := \psi_\ell + \hat{\varphi}, ~~\ell\in\{L\}. 
        \end{aligned}
    \end{equation}
\end{lemma}
\begin{IEEEproof}
The proof is provided in {\bf Appendix~\ref{app:proof_Wirtinger_derivative}}. 
\end{IEEEproof}

The aim of {\bf Lemma~\ref{lemma:wirtinger_der_vM-EM}} is to establish the connection between the vM-EM estimator $\hat{\varphi}$ and the noise vector ${\bm v}$. In the following, we denote the estimation error as 
\begin{equation}
    \Delta\varphi := \hat{\varphi} - \varphi. 
\end{equation}
Specifically, it is justified in {\bf Lemma~\ref{lemma:wirtinger_der_vM-EM}} that, the squared error $|\Delta\varphi|^2$ of the vM-EM estimator is intrinsically bounded by the noise energy $\|{\bm v}\|^2$, since the differential ${\rm d}\Delta\varphi$ of a general complex-valued function $\Delta\varphi: \mathbb{C}\to\mathbb{C}$ can be written as 
\begin{equation}
    {\rm d}\Delta\varphi = \langle \nabla_{\bm v}\hat{\varphi}, {\rm d}{\bm v}^*\rangle + \langle \nabla_{{\bm v}^*}\hat{\varphi}, {\rm d}{\bm v}\rangle, 
\end{equation}
where for real-valued function $\hat{\varphi}({\bm v}): \mathbb{C}^L \to [0,2\pi]\subset \mathbb{R}$ this differential relation reduces to 
\begin{equation}
    {\rm d}\Delta\varphi = 2\Re\langle\nabla_{\bm v}\hat{\varphi}, {\rm d}{\bm v}^*\rangle. 
    \label{eqn:differential_of_error}
\end{equation}
Thus, an integration inequality will hold to upper-bound the MSE of the estimator $\hat{\varphi}$, i.e., 
\begin{equation}
    \begin{aligned}
        |\Delta{\varphi}|^2 &= \left| \int {\rm d}\Delta\varphi \right|^2 \\
        &= 4 \left| \int_{0}^{1} \Re\langle\nabla_{{\bm v}'=t {\bm v}}\hat{\varphi}, {\bm v}\rangle{\rm d}t \right|^2 \\
        &\leq 4\|{\bm v}\|^2 \left(\int_{0}^{1} \|\nabla_{{\bm v}'=t{\bm v}}\hat{\varphi} \| {\rm d} t \right)^2. 
    \end{aligned}
    \label{ineq:gradient_bound}
\end{equation}
Before explaining this idea in detail, we first introduce some interesting results about some intermediate variables.

\begin{lemma}[Asymptotic invariants]\label{lemma:asymp_inv}
    Suppose $\alpha > \beta > 0$. If we define 
    \begin{equation}
        \begin{aligned}
            H_L & :=\frac{1}{L}\sum_{\ell=0}^{L-1} \frac{\sin^2(\theta_\ell)}{|\mu_\ell|^2}, \\
            G_L & := \frac{1}{L}\sum_{\ell=0}^{L-1} \frac{(\alpha+\beta \cos(\theta_\ell))(\beta+\alpha\cos(\theta_\ell))}{|\mu_\ell|^2}, \\
        \end{aligned}
    \end{equation}
    then the sequences $H_L$ and $G_L$ are intrinsically independent of the estimator $\hat{\varphi}\in[0,2\pi]$ as $L\to\infty$. Specifically, 
    \begin{equation}
        \begin{aligned}
            H &= H_{\infty} + \mathcal{O}(L^{-1}), \\
            G &= G_{\infty} + \mathcal{O}(L^{-1}), 
        \end{aligned}
    \end{equation}
    where 
    \begin{equation}
        \begin{aligned}
            H_{\infty} &=  \frac{1}{2\pi}\int_{0}^{2\pi} \frac{\sin^2\theta}{\alpha^2+\beta^2+2\alpha\beta\cos(\theta)}{\rm d}\theta = \frac{1}{2\alpha^2},  \\
            G_{\infty} &= \frac{1}{2\pi}\int_{0}^{2\pi} \frac{(\alpha+\beta\cos(\theta))(\beta+\alpha\cos(\theta))}{\alpha^2+\beta^2+2\alpha\beta\cos(\theta)} {\rm d}\theta = \frac{\beta}{2\alpha}.  \\ 
        \end{aligned}
    \end{equation}
\end{lemma}

{\it Proof Sketch.} The integral expressions for $H_\infty$ and $G_\infty$ as well as the asymptotic residuals $\mathcal{O}(L^{-1})$ can be obtained by applying the numerical trapezoidal integration formula~\cite{burden2015numerical} to the definition of the sequence $H_L$ and $G_L$. The integrals on $[0,2\pi]$ can be evaluated by the following replacements~\cite{martin1966complex}:
\begin{equation}
    \begin{aligned}
        & {\rm d}\theta  \to \frac{\d z}{\ri z}, & \int_0^{2\pi} \to \int_{|z|=1}, \\
        & \sin\theta  \to \frac{z-z^{-1}}{2\ri}, & \cos\theta \to \frac{z+z^{-1}}{2}, \\
    \end{aligned}
\end{equation}
and the applying the Residue Theorem to all poles inside the closed curve $|z|=1$. 

\begin{remark}
    These limiting expressions will play an important rule in obtaining the upper bound of the estimation error $|\Delta\varphi|$, since this invariance can asymptotically eliminate the dependence of the Wirtinger derivative ({\bf Lemma~\ref{lemma:wirtinger_der_vM-EM}}) on the unknown estimator $\hat{\varphi}$. 
\end{remark}

\begin{remark}
    In fact, we can also prove that 
    \begin{equation}
        \frac{\|{\bm x}\|^2}{L} = \frac{2\alpha^2-\beta^2/2}{4\alpha^2(\alpha^2-\beta^2)} + \mathcal{O}(L^{-1}),~~\alpha>\beta>0
    \end{equation}
    by the same numerical integration approximation technique and the Residue Theorem~\cite{martin1966complex}. 
\end{remark}

\begin{lemma}[ODE bound] \label{lemma:ODE-bound}
    Suppose for the input data sequence ${\bm s} = \sqrt{{\bm P}/A} = (s_0, \cdots, s_{L-1})\T$, the vM-EM algorithm converges, and the estimation error is denoted by $\Delta\varphi = \hat{\varphi} - \varphi$. Then, for any sufficiently large $L\geq L_0$, there exists some positive $\delta = \delta(L)>0$ and $C=C(L)>0$, such that for any noise vector ${\bm v}$ satisfying $\|{\bm v}\|/\sqrt{L} \leq \delta$, the estimation error is upper-bounded by
    \begin{equation}
        |\Delta\varphi|\leq C \frac{\|{\bm v}\|}{\sqrt{L}}. 
    \end{equation}
\end{lemma}
\begin{IEEEproof}
    The proof is provided in {\bf Appendix~\ref{app:proof_ODE_bound}}. 
\end{IEEEproof}
From the above {\bf Lemma~\ref{lemma:ODE-bound}}, we can directly prove {\bf Theorem~\ref{thm:asymp_perf_vM-EM}}. Choose $L\geq L_0$, and then let us compute the MMSE of the vM-EM estimator, i.e., 
\ifx\onecol\undefined
\begin{equation}
    \begin{aligned}
    \mathbb{E}[(\hat{\varphi}-\varphi)^2] &= \mathbb{E}[|\Delta\varphi|^2] \\
    &= \mathbb{E}\left[|\Delta\varphi|^2 \Bigg|\frac{\|{\bm v}\|}{\sqrt{L}} \leq \delta \right]\mathbb{P}\left[\frac{\|{\bm v}\|}{\sqrt{L}} \leq \delta \right] \\
    & + \mathbb{E}\left[|\Delta\varphi|^2 \Bigg|\frac{\|{\bm v}\|}{\sqrt{L}} > \delta \right]\mathbb{P}\left[\frac{\|{\bm v}\|}{\sqrt{L}}>\delta \right] \\
    & \leq C^2 \mathbb{E}\left[\frac{\|{\bm v}\|^2}{L}\right] + \pi^2 \mathbb{P}\left[\frac{\|{\bm v}\|}{\sqrt{L}}>\delta\right] \\
    & \overset{(a)}{\leq} C^2\sigma_v^2 + \pi^2 \frac{\mathbb{E}\left[\left(\frac{\|{\bm v}\|}{\sqrt{L}}\right)^r\right]}{\delta^r},
    \end{aligned}
\end{equation}
\else 
\begin{equation}
    \begin{aligned}
    \mathbb{E}[(\hat{\varphi}-\varphi)^2] &= \mathbb{E}[|\Delta\varphi|^2] \\
    &= \mathbb{E}\left[|\Delta\varphi|^2 \Bigg|\frac{\|{\bm v}\|}{\sqrt{L}} \leq \delta \right]\mathbb{P}\left[\frac{\|{\bm v}\|}{\sqrt{L}} \leq \delta \right] + \mathbb{E}\left[|\Delta\varphi|^2 \Bigg|\frac{\|{\bm v}\|}{\sqrt{L}} > \delta \right]\mathbb{P}\left[\frac{\|{\bm v}\|}{\sqrt{L}}>\delta \right] \\
    & \leq C^2 \mathbb{E}\left[\frac{\|{\bm v}\|^2}{L}\right] + \pi^2 \mathbb{P}\left[\frac{\|{\bm v}\|}{\sqrt{L}}>\delta\right] \\
    & \overset{(a)}{\leq} C^2\sigma_v^2 + \pi^2 \frac{\mathbb{E}\left[\left(\frac{\|{\bm v}\|}{\sqrt{L}}\right)^r\right]}{\delta^r},
    \end{aligned}
\end{equation}
\fi
where (a) comes from applying the Markov inequality, and $r>0$ can be arbitrarily chosen. Particularly, by setting $r=2$, we obtain 
\begin{equation}
    \mathbb{E}\left[(\hat{\varphi} - \varphi)\right] \leq \sigma_v^2(C^2+\pi^2/\delta^2) = \mathcal{O}(\bar{\gamma}^{-1}),  
\end{equation}
which completes the proof of {\bf Theorem~\ref{thm:asymp_perf_vM-EM}}. 

\begin{remark}
    The conclusion of {\bf Theorem~\ref{thm:asymp_perf_vM-EM}} guarantees that as $\bar{\gamma}\to\infty$, the MSE of the vM-EM estimator decays at a rate of at least $(\bar{\gamma})^{-1}$. This conclusion is verified in the following numerical simulation, where it is shown that the MSE curve of the proposed vM-EM algorithm has an asymptotic slope of $-1$ in the logarithmic coordinate. 
\end{remark}

}

\subsection{Expressions for the CRLB} \label{Precise CRLB}
In Section~\ref{Sensing RIS-Based Channel Estimation}, we have introduced three phase estimation methods to solve the probabilistic parameter estimation problem.
To analyze the performance limit of the proposed schemes, we derive the CRLB~\cite{casella2021statistical} of the estimation in \textbf{Lemma~\ref{lemma:precise CRLB}}.
    
    \begin{lemma}[Non-central Chi-Squared CRLB] \label{lemma:precise CRLB}\mbox{}\par
        Suppose the probabilistic model is specified by~\eqref{eqn:sensor power}, where $L$ observations $P[0],\cdots,P[L-1]$ are obtained. Then the CRLB of this $L$-point phase estimation problem is given by
        \begin{equation}
            \frac{1}{{\rm CRLB}(\varphi)} = K^2(\bar{\gamma})^2 \sum_{l=0}^{L-1}\sin^2(\psi_l+\varphi)\left(1/\gamma_l-g(\gamma_l)\right),
            \label{eqn:precise_CRLB}
        \end{equation}
        where $K=2\alpha\beta/(\alpha^2+\beta^2)$, $a=A\sigma_v^2$, $\gamma_l=\lambda_l/a=\left(\alpha^2+\beta^2+2\alpha\beta \cos(\psi_l+\varphi)\right)/\sigma_v^2$, and $\bar{\gamma}$ is the arithmetic mean of all the $\gamma_0,\cdots,\gamma_{L-1}$. The function $g(\gamma)$ is defined as 
        \ifx\onecol\undefined
            \begin{equation}\label{eqn:precise_g_function}
                \begin{aligned}
                    & g(\gamma)  = \int_{0}^{+\infty}\gamma t \, {\rm exp}(-\gamma(1+t))\, I_0\left(2\gamma\sqrt{t}\right) \\
                    & \left(1-R^2\left(2\gamma\sqrt{t}\right)\right){\rm d}t.
                \end{aligned}
            \end{equation}
        \else 
            \begin{equation}
                g(\gamma)  = \int_{0}^{+\infty}\gamma t \, {\rm exp}(-\gamma(1+t))\, I_0\left(2\gamma\sqrt{t}\right)\,\left(1-R^2\left(2\gamma\sqrt{t}\right)\right){\rm d}t.
                \label{eqn:precise_g_function}
            \end{equation}
        \fi
        The function $R(\cdot)$ is defined the same as in~\eqref{eqn:R function derivative property}.
    \end{lemma}
    \begin{IEEEproof}
        See {\bf Appendix~\ref{app:proof of precise CRLB}}. 
    \end{IEEEproof}

    From \textbf{Lemma~\ref{lemma:precise CRLB}}, we note that the CRLB of the phase estimation problem is completely determined by the two parameters $K$ and $\bar{\gamma}$, which are the interferential contrast and the average interferential SNR, respectively. 
    However, this exact CRLB expression is difficult to calculate due to the sophisticated evaluation of the integral in~\eqref{eqn:precise_g_function}.
    Thus, we will provide a simpler approximated CRLB in the following {\bf Theorem~\ref{thm:asymptotic CRLB}}. 
    
    \begin{theorem}[Asymptotic CRLB] \label{thm:asymptotic CRLB} \mbox{}\par
        Suppose the probabilistic model is specified by~\eqref{eqn:sensor power}. Then the approximated CRLB of this $L$-point phase estimation problem is given by 
        \begin{equation}
            \frac{1}{{\rm CRLB}(\varphi)}\approx K^2(\bar{\gamma})^2 \sum_{l=0}^{L-1}\sin^2(\psi_l+\varphi)(1/\gamma_l-\hat{g}(\gamma_l))^{+},
            \label{eqn:asymptotic CRLB}
        \end{equation}
        where $(x)^{+}$ represents ${x{\mathbbm{1}}_{\{x\geq 0\}}}$, and the definition of parameters are the same as in {\bf Lemma \ref{lemma:precise CRLB}}. The asymptotic approximation function $\hat{g}(\gamma)$ is defined as 
        \begin{equation}
            \hat{g}(\gamma) = \frac{1}{4} \sqrt{\frac{\pi}{\gamma}}e^{-\gamma/2}\left((1+1/\gamma)I_0(\gamma/2) + I_1(\gamma/2)\right).
            \label{eqn:definition g function}
        \end{equation}
    \end{theorem}
    \begin{IEEEproof}
        See {\bf Appendix \ref{Proof of Theorem 2}}. 
    \end{IEEEproof}
    
        The asymptotic expression of the CRLB still relies only on two parameters $K$ and $\bar{\gamma}$. But the calculation is much simpler compared to the exact expression. 
        The discussions of the properties of the exact and asymptotic expressions of CRLB are continued in Subsection~\ref{Properties of the CRLB}.
        
\subsection{Properties of the CRLB} \label{Properties of the CRLB}
\begin{figure}[!t]
        \centering
        \myincludegraphics{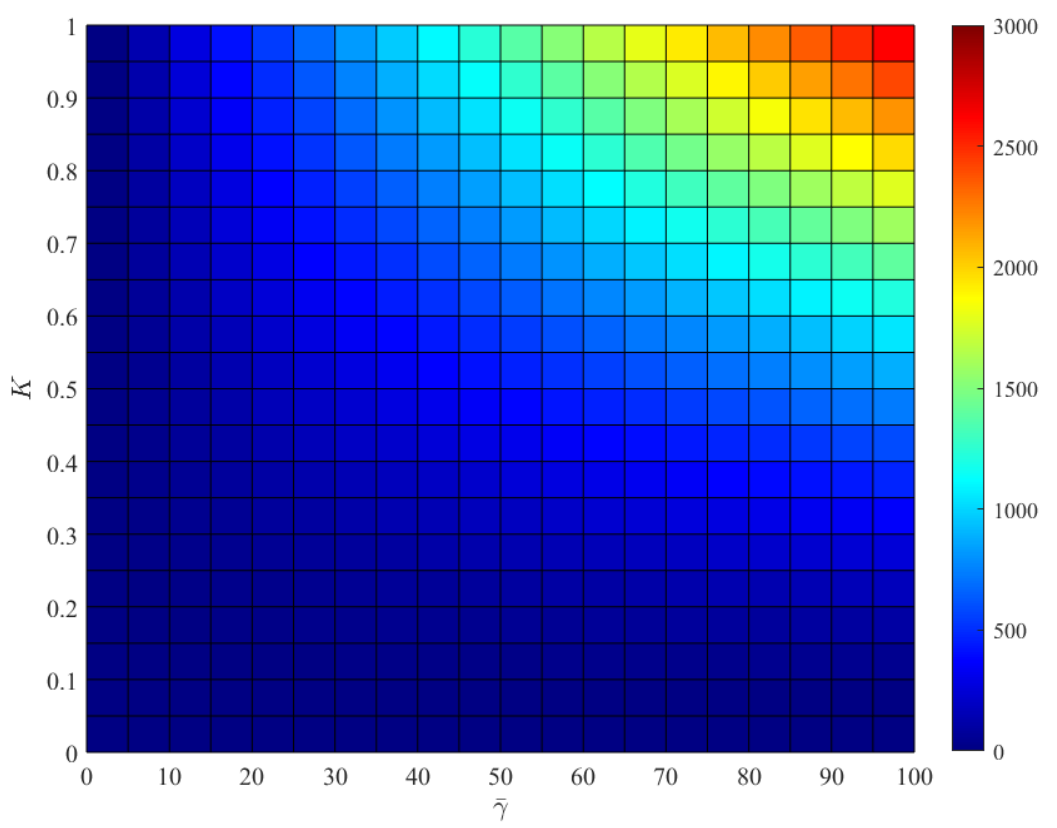}
        \caption{Precise $1/{\rm CRLB}(\varphi)$ as a two-variable function of $K$ and $\bar{\gamma}$, calculated from~\eqref{eqn:precise_CRLB}. The larger the value of the reciprocal CRLB, the more precise an unbiased estimator can be. }
        \label{fig:CRLB two variable}
    \end{figure}

    From the above analysis of CRLB, we can observe two properties of the phase estimation problem:
    \begin{enumerate}
        \item The CRLB relies almost only on the physical parameters $K$ and $\bar{\gamma}$.
        \item The CRLB is insensitive to the value $\varphi$.
    \end{enumerate}
    The first point can be concluded from the expressions, and the second point comes from the symmetry of the $\sin^2(\cdot)$ function when $L$ is large enough. 
    We have plotted $1/{\rm CRLB}(\varphi)$ as a two-variable function in Fig.~\ref{fig:CRLB two variable}. 
    The best prediction accuracy occurs when $\lvert K \rvert = 1$, i.e., the RIS-received signal from BS is as strong as that from the user. 
    Also, the larger value of the average interferential SNR $\bar{\gamma}$ also contributes to a more accurate phase estimation. 

We also perform error analysis for the use of asymptotic expansion technique, which is stated in the following {\bf Theorem~\ref{thm:CRLB_asym_error_analysis}}. 

    \begin{theorem}[Asymptotic Optimality of CRLB] \label{thm:CRLB_asym_error_analysis}\mbox{}\par
        The relative error $r$ of $1/\gamma - \hat{g}(\gamma)$, as an approximation of $1/\gamma - g(\gamma)$, decreases at a rate that is inverse proportional to the interferential SNR $\gamma$, i.e.,
        \begin{equation}
            r=\frac{\left|\hat{g}(\gamma)-g(\gamma)\right|}{1/\gamma - g(\gamma)} = \mathcal{O}\left(\frac{1}{\gamma}\right).
        \end{equation}
    \end{theorem}
    \begin{IEEEproof}
        See {\bf Appendix~\ref{Proof of Theorem 3}}. 
    \end{IEEEproof}

From {\bf Theorem~\ref{thm:CRLB_asym_error_analysis}}, we can conclude that our derived expression of approximated CRLB is asymptotically optimal.

\section{Simulation Results}
\label{Simulation Results}
In this section, we will present our simulation results. In Subsection~\ref{Phase Estimation Algorithms}, we will show the performance comparison of the phase estimation algorithms. In Subsection~\ref{Achievable Spectral Efficiency under IRF}, we will show the achievable rate comparison of our \ac{IRF} method and other RIS-aided channel estimation and beamforming algorithms. 

\subsection{Phase Estimation Algorithms} \label{Phase Estimation Algorithms}
    In Section~\ref{Sensing RIS-Based Channel Estimation}, we have already introduced three phase estimation algorithms: DFT, Newton-ML and vM-EM. Now we compare the performance of these algorithms together with the CRLB which has been derived in Section~\ref{Performance Analysis}. 
    \begin{figure}[!t]
        \centering
        \myincludegraphics{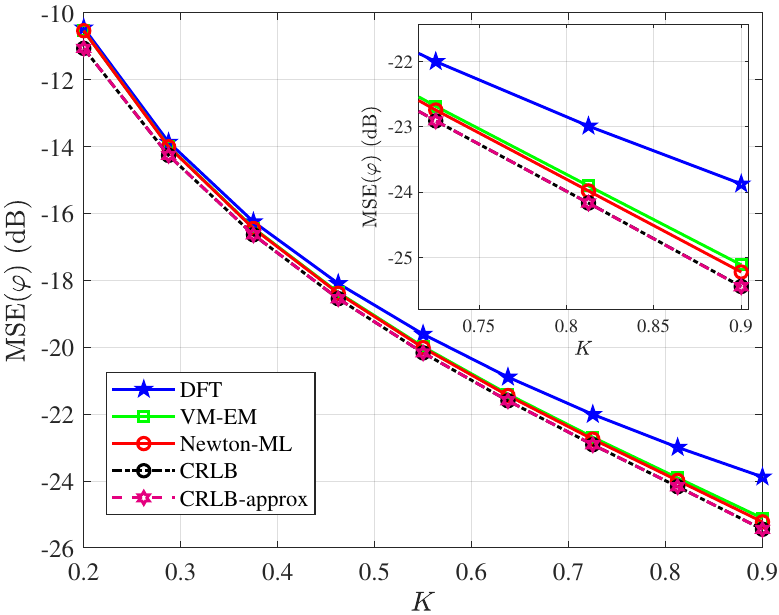}
        \caption{Performance comparison of phase estimation algorithms. $x$-axis represents the interferential contrast $K$; $y$-axis represents the MSE of the estimators. The vM-EM algorithm outperforms naive DFT by at least $1\,{\rm dB}$ in high-$K$ regions. }
        \label{fig:phase estimation_K}
    \end{figure}
    \begin{figure}[!t]
        \centering
        \myincludegraphics{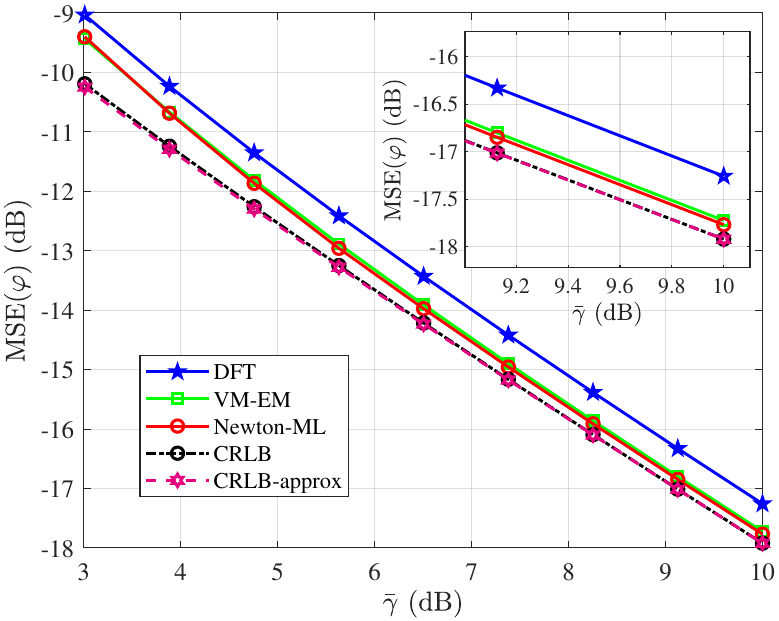}
        \caption{Performance comparison of phase estimation algorithms. $x$-axis represents the average interferential SNR $\bar{\gamma}$; $y$-axis represents the MSE of the estimators. The interferential contrast $K=0.6$. }
        \label{fig:phase estimation_gamma}
    \end{figure}
    In all the simulations for phase estimation algorithms, the amplification factor $A$ is fixed to be $1$, the number of power samples\footnote{In practical OFDM systems, the number of signal samples within the duration of an OFDM symbol is usually $\sim 2048$. Considering that the power sensor is usually slower than the baseband ADC, it is reasonable to assume a 32-times slower sampling rate of the power sensor. Thus, $L=64$ is a reasonable number of acquired power samples during one time slot.} is fixed to $L=2^6=64$, and $\sigma_\zeta=0.05$; 
    The interferential SNR $\bar{\gamma}=20$ in Fig.~\ref{fig:phase estimation_K}. 
    The CRLB and CRLB-approx curves are calculated from \eqref{eqn:precise_CRLB} and \eqref{eqn:asymptotic CRLB}, respectively. 
    For Newton-ML algorithm, the Newton iteration is performed 4 times according to \eqref{eqn:Newton-ML renewal formula}. 
    For vM-EM algorithm, the iteration number is also fixed to 4. The true value of random variable $\varphi$ under estimation is drawn from a uniform distribution on $[0,2\pi]$. 
    In the simulations in Fig.~\ref{fig:phase estimation_gamma}, all the simulation parameters, except for $K$ and $\bar{\gamma}$, are the same with that of Fig.~\ref{fig:phase estimation_K}.

    It can be concluded from Fig.~\ref{fig:phase estimation_K} and Fig.~\ref{fig:phase estimation_gamma} that, the vM-EM algorithm has comparable performance with the Newton-ML algorithm, but the computational cost is significantly lower, since {\bf Algorithm~\ref{alg:vM-EM}} does not require the evaluation of the complicated modified Bessel functions. 
    
    It is shown in Fig.~\ref{fig:phase estimation_gamma} that the vM-EM algorithm achieves an asymptotic error decay of $\mathcal{O}(\bar{\gamma}^{-1})$, which coincides with the statement in {\bf Theorem~\ref{thm:asymp_perf_vM-EM}}. Furthermore, both the Newton-ML and vM-EM algorithms are close to the CRLB, and both of them outperform the simple DFT algorithm. The CRLB approximation~\eqref{eqn:asymptotic CRLB} is also satisfactory under a wide range of $K$ and $\bar{\gamma}$. However, the DFT estimator in Fig.~\ref{fig:phase estimation_gamma} exhibits a near-constant performance gap toward the CRLB in the high-SNR region, while the Newton-ML and vM-EM estimators are able to bridge this gap and finally approaches the CRLB as $\bar{\gamma}\to\infty$.

\subsection{Spectral Efficiency with IRF-based CSI Acquisition} \label{Achievable Spectral Efficiency under IRF}

    All our simulation data are acquired under $f_c = 3.5\,{\rm GHz}$, $P_{\rm max}'=300 \,{\rm mW}$, $n_0=-174\,{\rm dBm/Hz}$, subcarrier bandwidth ${\rm BW} = 180\,{\rm kHz}$, thermal noise at the receiver $\sigma^2 = {\rm BW}\times n_0$, $\sigma_v^2=100\,{\rm MHz}\times n_0 F_p$, where $F_p=10\,{\rm dB}$ is the noise factor of the power sensor. 
    Both the BS and the RIS are equipped with $\lambda/2$-spaced uniform planar arrays (UPAs), while the user has a single antenna. 
    The BS is located at a distance from the RIS that is drawn from a uniform distribution between $20\,{\rm m}$ and $100\,{\rm m}$, and the user appears uniformly within a distance ranging from $10\,{\rm m}$ to $100\,{\rm m}$.
    The size of the RIS is set to be $N=20\times 10$, and that of the BS antenna is $4\times 2$. The detailed simulation settings about the pilot energy and data energy~\eqref{eqn:transmit_power} are listed in {\bf Table~\ref{tab:sim_parameters}}.
    
    \begin{table}[t] 
        \centering
        \begin{threeparttable}
            \caption{Fair Pilot Overhead Comparison of Different CSI Acquisition Methods} \label{tab:sim_parameters}
            \begin{tabular}{l|c|c|c|c|c}
                \toprule
                CSI acquisition method  & $N_p$     & $N_d$ & $B$                   & $E_p$                 &  $E_d$            \\ 
                \hline 
                LMMSE $\bm H$           & 200       & 800   & \multirow{3}*{1000}   & \multirow{3}*{50}     & \multirow{3}*{950}\\
                \cline{1-3}
                MMSE $\bm f$            & 25        & 975   & ~                     & ~                     & ~                 \\
                \cline{1-3}
                Proposed IRF            & 3         & 997   & ~                     & ~                     & ~                 \\ 
                \bottomrule
            \end{tabular}
        \end{threeparttable}
    \end{table}
    
    \begin{figure}[!t]
        \centering 
        \myincludegraphics{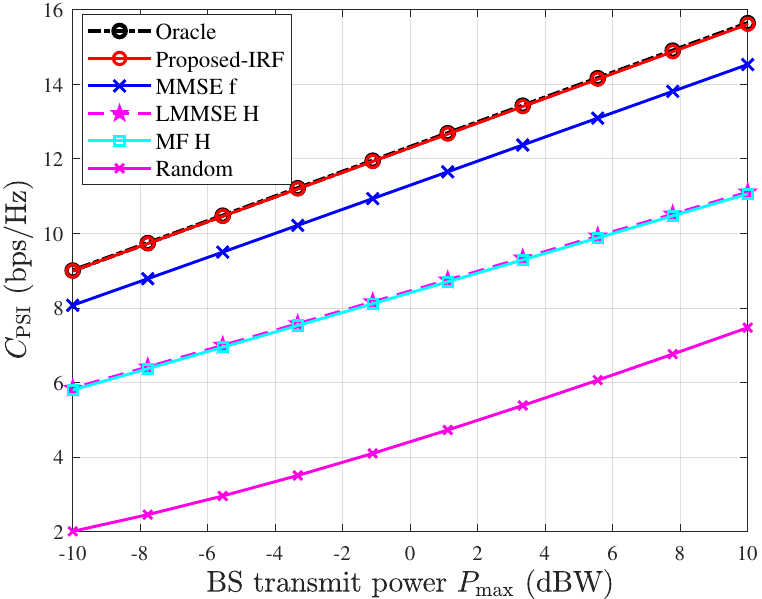}
        \caption{Performance curve of the ergodic spectral efficiency against the BS transmitted power $P_{\rm max}$.}
        \label{fig:rate}
    \end{figure}
    \red{
    Moreover, for the IRF methods, the BS beamformer $\bm w$ is chosen to be the right singular vector that corresponds to the most significant singular value of the BS-RIS channel $\bm G$, which maximizes the total signal energy received by the RIS from the BS, i.e., 
    \begin{equation}
        {\bm w} = \underset{\left\|{\bm w}\right\|\leq P_{\rm max}}{\rm argmax}\left\|{\bm G}{\bm w} \right\|^2,
    \end{equation}
    and the user's transmit scaler is set to the maximum allowed power $w'=\sqrt{P'_{\rm max}}$ for all the CSI acquisition schemes to ensure a high SNR. 
    For the MMSE/LMMSE algorithms, after the channel estimation procedure, the beamforming step is fulfilled by alternately updating the BS beamformer $\bm w$ and the RIS phase-shift matrix $\bm\Theta = {\rm diag}({\bm \theta})$ according to~\eqref{eqn:iter_beamforming}. 
    The initial values ${\bm w}^{(0)}$ and ${\bm\theta}^{(0)}$ are generated randomly, and are subject to their corresponding power constraint and unit-modulus constraint, respectively. In our simulations, the iteration is performed for at most $T=10$ times. 
    
    The proposed \ac{IRF} channel estimation and beamforming method are compared with the conventional MMSE-$\bm f$, LMMSE-$\bm H$, and other idealized settings in Fig.~\ref{fig:rate}. 
    In the simulation of the IRF method, the vM-EM phase estimation algorithm is utilized with the assumption that the phases of the BS-RIS \ac{LOS} path are known. 
    The ``Oracle'' (black dashed line) assumes perfectly-known CSI with iterative beamforming~\eqref{eqn:iter_beamforming}. 
    In the LMMSE-$\bm H$ method (pink dashed line), we first utilize~\eqref{eqn:LMMSE} to obtain estimates for the cascaded channel~\cite{kundu2021channel,wei2021channel}, and then perform iterative RIS beamforming~\eqref{eqn:iter_beamforming} based on the estimated channel.
    The MMSE-$\bm f$ method (blue line) is calculated by~\eqref{eqn:MMSE-f-estimator} before applying the beamforming algorithm~\eqref{eqn:iter_beamforming}. 
    The matched filter (MF) scheme (cyan line) is obtained by assuming $\sigma_z^2\to\infty$ in~\eqref{eqn:LMMSE}. 
    In addition, we also consider the randomly-phased RIS scheme as a benchmark.  

    Fig.~\ref{fig:rate} shows the performance curve of the ergodic spectral efficiency $C_{\rm PSI}$ against the BS transmitted power. The ergodic spectral efficiency is defined as 
    \begin{equation}
        C_{\rm PSI} = \frac{N_d}{B} \mathbb{E}\left[\log\left( 1+\gamma_{\rm user} \right)\right], \label{eqn:capacity_def}
    \end{equation} 
    which is normalized by the pilot overhead, and the subscript ${\rm PSI}$ represents perfect side information~\cite{lapidoth2002fading}. The achievable user SNR is defined as 
    \begin{equation}
        \gamma_{\rm user} = \frac{P_{{\rm BS}, d} |{\bm f}\H {\bm \Theta} {\bm G}{\bm w}|^2}{\sigma_z^2}.
    \end{equation}
    From Fig.~\ref{fig:rate}, we can conclude that the proposed IRF method with vM-EM phase estimation is nearly optimal even with extremely low pilot overhead. The proposed IRF method is even close to the oracle scheme, and achieves satisfactory performance throughout a wide range of the BS transmitted power. 
    }

\section{Conclusion}
\label{Conclusion}
    \red{In this paper, we have introduced a dimension-independent CSI acquisition method for sensing RIS-assisted MISO wireless communication systems. 
    Combined with our proposed vM-EM phase estimation algorithm, the pilot overhead of our CSI acquisition method is made independent of the number of RIS elements with low computational cost, enabling the implementation of extremely large-scale RISs to achieve significant beamforming gain. 
    Theoretical analysis have demonstrated the asymptotic optimality of the proposed vM-EM algorithm, which is further supported by the CRLB analysis.  Simulation results have also verified the near-optimality of our vM-EM algorithm. }
    Furthermore, due to the elementwise independent property of our \ac{IRF}-based CSI acquisition method, the near-field effect cannot corrupt the precision of the CSI. 
    Also, due to the simultaneous signaling protocol, the ``multiplicative fading'' effect of RIS~\cite{zhang2021active,liu2021active} during channel estimation is automatically avoided. 
    Thus, the \ac{IRF} method has promising applications to high-frequency large-scale systems. 
    
    \red{For future work, the spatial interferential fringes on the sensing RIS may be exploited to recover the CSI at higher precision, and the data obtained by the power sensors may be utilized to perform joint channel estimation and beamforming with sparse assumptions on the channel. 
    In addition, different interferential frequencies can be assigned to different users to perform multi-user CSI acquisition simultaneously, but the waveforms should be re-designed to avoid interference among users. }
    Furthermore, when equipped with a sensing RIS, traditional MIMO systems can also benefit from the additional CSI provided by the phase estimation methods based on the IRF.  

\appendices
\section{Proof of \textbf{Lemma 1}}
\label{Proof of Lemma 1}
        Denote $z=z_r + {\ri} z_i$; then the posterior density $p(\theta | z) \propto p(\theta)p(z|\theta)$ can be expressed as
        \AutoColumn{
            \begin{equation}
                \begin{aligned}
                    p(\theta|z) & \propto \exp(\kappa \cos(\theta - \mu))\exp\left(-\frac{1}{\sigma^2}\left(\left(z_r - \cos\theta\right)^2 + \left(z_i - \sin\theta\right)^2\right)\right) \\
                    & \propto \exp\left( \kappa \cos(\theta - \mu) + \frac{2}{\sigma^2}(z_r \cos\theta + z_i\sin\theta) \right) \\
                    & \propto \exp\left( \re\left[ e^{{\ri} \theta}\left(\kappa e^{-{\ri}\mu} + \frac{1}{\sigma^2/2} z^*\right)\right] \right).
                \end{aligned}
            \end{equation}
        }{
            \begin{equation}
                \begin{aligned}
                    p(\theta|z) & \propto \exp(\kappa \cos(\theta - \mu)) \\
                    &~~~\times \exp\left(-\frac{1}{\sigma^2}\left(\left(z_r - \cos\theta\right)^2 + \left(z_i - \sin\theta\right)^2\right)\right) \\
                    & \propto \exp\left( \kappa \cos(\theta - \mu) + \frac{2}{\sigma^2}(z_r \cos\theta + z_i\sin\theta) \right) \\
                    & \propto \exp\left( \re\left[ e^{{\ri} \theta}\left(\kappa e^{-{\ri}\mu} + \frac{1}{\sigma^2/2} z^*\right)\right] \right).
                \end{aligned}
            \end{equation}
        }

        Since the density of the von Mises distribution $\VM(\mu, \kappa)$ can also be expressed as $p(\theta) \propto \exp\left( \re\left[ e^{{\ri} \theta}(\kappa e^{{\ri} \mu})^* \right] \right)$, we can also parameterize the von Mises distribution by a single complex parameter $\kappa e^{{\ri}\mu}$. Thus, the above $p(\theta|z)$ is a von Mises density with parameter $\kappa'e^{{\ri}\mu'}$, satisfying $\kappa' e^{{\ri}\mu'} = \kappa e^{{\ri}\mu} +2z/\sigma^2$. This completes the proof.

\section{Proof of \textbf{Lemma~\ref{lemma:precise CRLB}}}\label{app:proof of precise CRLB}
    According to the definition of the CRLB, taking the negative expectation of \eqref{Second Derivative Likelihood} yields the reciprocal CRLB of the estimators for $\varphi$. Note that in \eqref{Second Derivative Likelihood}, there are three types of expectations to be evaluated: $\mathbb{E}\left[P[l]\right]$, $\mathbb{E}\left[R(z_l)\sqrt{P[l]}\right]$, and $\mathbb{E}\left[(1-R^2(z_l))P[l]\right]$. The expectation of $P[l]$ can be directly evaluated from \eqref{eqn:sensor power} by the linearity of the expectation operation:
    \begin{equation}
        \mathbb{E}\left[P[l]\right] = A\sigma_v^2 + \lambda_l.
        \label{eqn:expectation of P_l}
    \end{equation}
    The expectation $\mathbb{E}\left[R(z_l) \sqrt{P[l]}\right]$ can be evaluated by calculating the derivative w.r.t $\lambda_l$ on both sides of the identity $\mathbb{E}_{\lambda_l, A\sigma_v^2}[e^{\lambda_l/(A\sigma_v^2)}]=e^{\lambda_l/(A\sigma_v^2)}$:
    \AutoColumn{
        \begin{equation}
            \begin{aligned}
                \mathbb{E}\left[R(z_l) \sqrt{P[l]}\right] & = e^{-\lambda_l/(A\sigma_v^2)}\int_{0}^{+\infty}{\frac{1}{A\sigma_v^2}\exp\left(-\frac{x}{A\sigma_v^2}\right)I_1\left(\frac{\sqrt{\lambda_l x}}{A\sigma_v^2/2}\right)\sqrt{x}{\rm d}x} \\
                & = e^{-\lambda_l/(A\sigma_v^2)} A\sigma_v^2 \sqrt{\lambda_l} \frac{\partial}{\partial\lambda_l} \int_{0}^{+\infty}{\frac{1}{A\sigma_v^2}\exp\left(-\frac{x}{A\sigma_v^2}\right)I_0\left(\frac{\sqrt{\lambda_l x}}{A\sigma_v^2/2}\right) {\rm d}x}\\
                & = \sqrt{\lambda_l}.\\
            \end{aligned}
            \label{eqn:expectation_second_kind_1}
        \end{equation}
    }{
        \begin{equation}
            \begin{aligned}
                \mathbb{E}\left[R(z_l) \sqrt{P[l]}\right] & = e^{-\lambda_l/(A\sigma_v^2)}\times \int_{0}^{+\infty} \\
                & {\frac{1}{A\sigma_v^2}\exp\left(-\frac{x}{A\sigma_v^2}\right)I_1\left(\frac{\sqrt{\lambda_l x}}{A\sigma_v^2/2}\right)\sqrt{x}{\rm d}x} \\
                & = e^{-\lambda_l/(A\sigma_v^2)} A\sigma_v^2 \sqrt{\lambda_l} \times \\
                &\frac{\partial}{\partial\lambda_l} \int_{0}^{+\infty}{\frac{1}{A\sigma_v^2}\exp\left(-\frac{x}{A\sigma_v^2}\right)I_0\left(\frac{\sqrt{\lambda_l x}}{A\sigma_v^2/2}\right) {\rm d}x}\\
                & = \sqrt{\lambda_l}.\\
            \end{aligned}
            \label{eqn:expectation_second_kind_1}
        \end{equation}
    }
    
    According to the definition of $z_l$, the expectation $\mathbb{E}\left[R(z_l)P[l]/z_l\right]$ is of the same form as the expectation $\mathbb{E}\left[R(z_l) \sqrt{P[l]}\right]$. Thus, the expectation result is
    \begin{equation}
        \mathbb{E}\left[\frac{R(z_l)}{z_l} P[l]\right] = \frac{A\sigma_v^2}{2}.
        \label{eqn:expectation_second_kind_2}
    \end{equation}
    As for the third kind of expectation $\mathbb{E}\left[(1-R^2(z_l))\frac{P[l]}{\lambda_l}\right]$, we first define 
    \begin{equation}
        g(\lambda_l, a) = \mathbb{E}\left[(1-R^2(z_l))\frac{P[l]}{\lambda_l}\right].
        \label{eqn:two-variable g function}
    \end{equation}
    Let $x=\lambda_l t$, and write down the integral expression of \eqref{eqn:two-variable g function}, we obtain 
    \AutoColumn{
        \begin{equation}
            \begin{aligned}
            g(\lambda_l, a) & = \int_{0}^{+\infty} t(1-R^2(2\gamma_l\sqrt{t})) \frac{\lambda_l}{a} \exp\left(\frac{\lambda_l + \lambda_l t}{a}\right) I_0\left(2\gamma_l\sqrt{t}\right){\rm d}t \\
            & = \int_{0}^{+\infty}\gamma_l t \, \exp(-\gamma_l(1+t))\, I_0\left(2\gamma_l\sqrt{t}\right)\,\left(1-R^2\left(2\gamma_l\sqrt{t}\right)\right){\rm d}t \\
            & = g(\gamma_l).
            \end{aligned}
            \label{eqn:proof precise g function}
        \end{equation}
    }{
        \begin{equation}
            \begin{aligned}
            g(\lambda_l, a) & = \int_{0}^{+\infty} t(1-R^2(2\gamma_l\sqrt{t})) \frac{\lambda_l}{a} \exp\left(\frac{\lambda_l + \lambda_l t}{a}\right) \\
            & \times I_0\left(2\gamma_l\sqrt{t}\right){\rm d}t \\
            & = \int_{0}^{+\infty}\gamma_l t \, \exp(-\gamma_l(1+t))\, I_0\left(2\gamma_l\sqrt{t}\right)\\
            &\times \left(1-R^2\left(2\gamma_l\sqrt{t}\right)\right){\rm d}t \\
            & = g(\gamma_l).
            \end{aligned}
            \label{eqn:proof precise g function}
        \end{equation}
    }
    
    Combining the above equations of expectations \eqref{eqn:expectation of P_l}, \eqref{eqn:expectation_second_kind_1}, \eqref{eqn:expectation_second_kind_2} and \eqref{eqn:proof precise g function}, we obtain
    \AutoColumn{
        \begin{equation}
            \begin{aligned}
            \frac{1}{{\rm CRLB}(\varphi)} 
            & \overset{(a)}{=}  -\frac{4\alpha^2\beta^2}{\sigma_v^4}\sum_{l=0}^{L-1}{ \sin^2(\psi(t_l)+\varphi) \left(\mathbb{E}\left[(1-R^2(z_l))\frac{P[l]}{\lambda_l}\right] - \frac{a}{\lambda_l}\right) } \\
            & \overset{(b)}{=} \frac{4\alpha^2\beta^2}{\sigma_v^4}\sum_{l=0}^{L-1}{ \sin^2(\psi(t_l)+\varphi) \left(\frac{a}{\lambda_l}-g(\lambda_l/a)\right)},
            \end{aligned}
            \label{eqn:CRLB}
        \end{equation}
    }{
        \begin{equation}
            \begin{aligned}
            \frac{1}{{\rm CRLB}(\varphi)} 
            & \overset{(a)}{=}  -\frac{4\alpha^2\beta^2}{\sigma_v^4}\sum_{l=0}^{L-1} \sin^2(\psi(t_l)+\varphi) \\
            & \times \left(\mathbb{E}\left[(1-R^2(z_l))\frac{P[l]}{\lambda_l}\right] - \frac{a}{\lambda_l}\right)  \\
            & \overset{(b)}{=} \frac{4\alpha^2\beta^2}{\sigma_v^4}\sum_{l=0}^{L-1}{ \sin^2(\psi(t_l)+\varphi) \left(\frac{a}{\lambda_l}-g(\lambda_l/a)\right)},
            \end{aligned}
            \label{eqn:CRLB}
        \end{equation}
    }

    where step $(a)$ comes from substituting these three types of expectation into \eqref{Second Derivative Likelihood}, and step $(b)$ comes from replacing the trickiest expectation by the definition of the function $g(\cdot)$ in \eqref{eqn:precise_g_function}.  
    
    Note that the precise value of the CRLB can be determined by the exact  single-variable $g$ function \eqref{eqn:precise_g_function}, whose variable $\gamma_l = \lambda_l / a$ is the interferential SNR. After further derivations, it can be observed from the expressions that the CRLB only relies on two intrinsic physical parameters: the interferential SNR $\gamma_l$, and the interferential contrast $K$ \cite{louradour1993interference}, which coincides with the physical intuition. Imitating the notations in optics, we define the parameter $K$ to be  
    \begin{equation}
        K = \frac{I_M - I_m}{I_M + I_m} = \frac{2\alpha\beta}{\alpha^2+\beta^2},
    \end{equation}
    and $K$ automatically satisfies $-1\leq K\leq 1$. Define the average interferential SNR $\bar{\gamma}$ to be the arithmetic average of $\gamma_l, 0\leq l < L$:
    \begin{equation}
        \bar{\gamma} = \frac{1}{L}\sum_{l=0}^{L-1}{\gamma_l} = \frac{\alpha^2+\beta^2}{\sigma_v^2}.
    \end{equation}
    Thus, the CRLB can be expressed as 
    \begin{equation}
        \frac{1}{{\rm CRLB}(\varphi)} = K^2(\bar{\gamma})^2 \sum_{l=0}^{L-1}{ \sin^2(\psi_l+\varphi) \left(1/\gamma_l-g(\gamma_l)\right) }
    \end{equation}
    where the values $\gamma_l=\bar{\gamma}\left(1+K\cos(\psi_l+\varphi)\right)$ are jointly determined by both the average interferential SNR $\bar{\gamma}$ and the interferential contrast $K$. This completes the proof.

\section{Proof of \textbf{Theorem~\ref{thm:asymptotic CRLB}}}
\label{Proof of Theorem 2}
    Since an exact expression of $g(\gamma)$ in~\eqref{eqn:precise_g_function} is difficult to calculate, we evaluate it approximately by utilizing the asymptotic expansion $x(1-R^2(2\sqrt{x})) \approx \sqrt{x}/2$~\cite{silverman1972special}. According to the definition, evaluating $g(\gamma)$ is the same as evaluating $\mathbb{E}[(1-R^2(z_l))P[l]/\lambda_l]$. 

    In order to evaluate the expectation $\mathbb{E}[(1-R^2(z_l))P[l]/\lambda_l]$, we first introduce some preliminaries about the noncentral chi distribution $\nc_{\chi_k}(\lambda)$ with noncentrality parameter $\lambda>0$. The distribution $\nc_{\chi_k}(\lambda)$ is the law of the length (2-norm) of a $k$-dimensional standard normal distribution $\mathcal{N}({\bm \mu}, {\bm I}_k)$, with $\lambda = \Vert {\bm \mu} \Vert_2$. Specifically, we are interested in the case where $k=2$, since this is the case of the 2-dimensional complex plane. For  $k=2$, let $Y \sim \nc_{\chi_2}(m)$, then we have \cite{park1961moments}
    \begin{equation}
        \mathbb{E}\left[Y\right] = \sqrt{\frac{\pi}{2}}L_{1/2}(-m^2/2),
        \label{eqn:noncentral chi mean}
    \end{equation}
    where $L_{1/2}$ denotes the generalized Laguerre function  of order $1/2$. The function $L_{1/2}(x)$ has explicit expression 
    \begin{equation}
        L_{1/2}(x) = e^{x/2}\left[(1-x)I_0\left(-\frac{x}{2}\right)-xI_1\left(-\frac{x}{2}\right) \right].
        \label{eqn:Laguerre half order}
    \end{equation}
    Recall that the asymptotic expansion $x(1-R^2(2\sqrt{x})) \sim \sqrt{x}/2$ holds for large $x$,
    and the random variable $P[l]$ obeys a non-central chi-squared distribution which can be equivalently expressed as 
    \AutoColumn{
        \begin{equation}
            P[l] \sim A\left| \CN\left( (\alpha + \beta e^{{\ri} (\psi_l + \varphi)}), \sigma_v^2 \right)\right|^2
            \sim \frac{a}{2} \left| \CN\left((\alpha + \beta e^{{\ri} (\psi_l + \varphi)})/(\sigma_v / \sqrt{2}), 2\right)\right|^2.
        \end{equation}
    }{
        \begin{equation}
            \begin{aligned}
            P[l] &\sim A\left| \CN\left( (\alpha + \beta e^{{\ri} (\psi_l + \varphi)}), \sigma_v^2 \right)\right|^2 \\
            & \sim \frac{a}{2} \left| \CN\left((\alpha + \beta e^{{\ri} (\psi_l + \varphi)})/(\sigma_v / \sqrt{2}), 2\right)\right|^2.
            \end{aligned}
        \end{equation}
    }

    Thus, we obtain
    \begin{equation}
        \begin{aligned}
        \mathbb{E}[(1-R^2(z_l))P[l]/\lambda_l] & = \mathbb{E}\left[\left(1-R^2\left(\frac{\sqrt{\lambda_l P[l]}}{a/2}\right)\right)\frac{P[l]}{\lambda_l}\right] \\
        & \approx \left(\frac{a}{\lambda_l}\right)^2 \mathbb{E}\left[ \sqrt{\frac{\lambda_l P[l]}{a^2}}/2 \right]\\
        & = \frac{1}{2}\frac{\sqrt{\lambda_l}}{a}\left(\frac{a}{\lambda_l}\right)^2 \sqrt{\frac{a}{2}}\mathbb{E}\left[\nc_{\chi_2}(m)\right],\\
        \end{aligned}
        \label{eqn:approx evaluation}
    \end{equation}
    where $m=\lvert\alpha + \beta e^{\ri (\psi_l + \varphi)}\rvert/(\sigma_v / \sqrt{2}) = \sqrt{2\lambda_l/a} = \sqrt{2\gamma_l}$, and thus $m^2/2=\gamma_l$. Plugging \eqref{eqn:noncentral chi mean} and \eqref{eqn:Laguerre half order} into \eqref{eqn:approx evaluation}, we obtain the final expression 
    \AutoColumn{
        \begin{equation}
            \begin{aligned}
            \mathbb{E}[(1-R^2(z_l))P[l]/\lambda_l] & \approx \frac{1}{2}\frac{\sqrt{\lambda_l}}{a}\left(\frac{a}{\lambda_l}\right)^2 \sqrt{\frac{a}{2}} \sqrt{\frac{\pi}{2}} e^{-\gamma_l/2}\left[(1+\gamma_l)I_0(\gamma_l/2)+\gamma_l I_1(\gamma_l/2)\right] \\
            & = \frac{1}{4}\sqrt{\frac{\pi}{\gamma_l}}e^{-\gamma_l/2}\left[(1+\gamma_l^{-1})I_0(\gamma_l/2)+ I_1(\gamma_l/2)\right] \\
            & := \hat{g}(\gamma_l).
            \end{aligned}
            \label{eqn:approx evaluation result}
        \end{equation}
    }{
        \begin{equation}
            \begin{aligned}
            & \mathbb{E}[(1-R^2(z_l))P[l]/\lambda_l] \\
            & \approx \frac{1}{2}\frac{\sqrt{\lambda_l}}{a}\left(\frac{a}{\lambda_l}\right)^2 \sqrt{\frac{a\pi}{4}} e^{-\gamma_l/2}\left[(1+\gamma_l)I_0(\gamma_l/2)+\gamma_l I_1(\gamma_l/2)\right] \\
            & = \frac{1}{4}\sqrt{\frac{\pi}{\gamma_l}}e^{-\gamma_l/2}\left[(1+\gamma_l^{-1})I_0(\gamma_l/2)+ I_1(\gamma_l/2)\right] \\
            & := \hat{g}(\gamma_l).
            \end{aligned}
            \label{eqn:approx evaluation result}
        \end{equation}
    }

    Finally, substituting the approximation \eqref{eqn:approx evaluation result} into the exact CRLB \eqref{eqn:precise_CRLB} yields the conclusion \eqref{eqn:asymptotic CRLB}, which completes the proof. Note that the operation $(x)^+$ in \eqref{eqn:asymptotic CRLB} ensures that each term of the CRLB is non-negative. 

\section{Proof of \textbf{Theorem~\ref{thm:CRLB_asym_error_analysis}}}
\label{Proof of Theorem 3}
    \begin{figure}[t]
        \centering
        \myincludegraphics{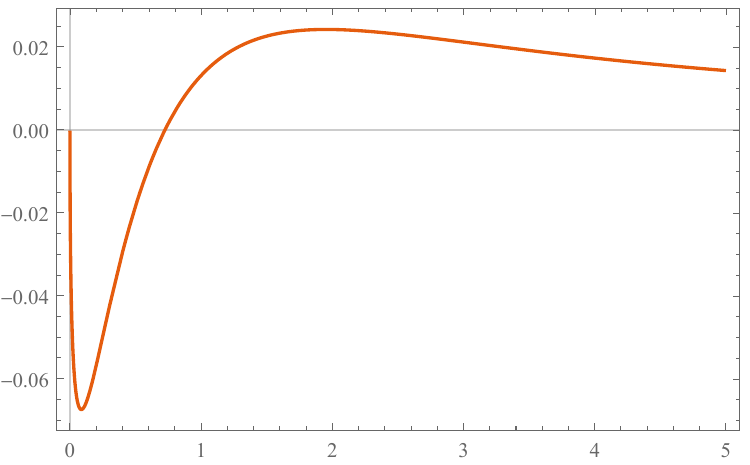}
        \caption{The curve $x(1-R^2(2\sqrt{x}))-\sqrt{x}/2$, when $0\leq x\leq 5$. We can see from the curve that the approximation error $\delta \leq 0.07$.}
        \label{fig:asymptotic_expansion}
    \end{figure}
    Since the only imprecise step of the above derivation is the asymptotic expansion, the approximation error of the expectation can be upper-bounded by the asymptotic expansion error. Assume that the asymptotic expansion error does not exceed $\delta$, i.e.,
    \begin{equation}
        \left\vert x\left(1-R^2(2\sqrt{x})\right)-\sqrt{x}/2 \right\vert \leq \delta.
        \label{eqn:asymptotic error}
    \end{equation}
    Then, by error analysis on \eqref{eqn:approx evaluation}, we have 
    \begin{equation}
        \left| g(\gamma) - \hat{g}(\gamma)\right| \leq \frac{1}{\gamma^2} \delta. 
    \end{equation}
    It can be numerically confirmed from Fig. \ref{fig:asymptotic_expansion} that $\delta \leq 0.07$, and that as $x\to \infty$, the approximation error tends to zero. Thus, \eqref{eqn:approx evaluation result} is a nearly perfect approximation when the interferential SNR $\gamma$ is large. To be more precise, at $\gamma \to +\infty$, the function $\hat{g}(\gamma)$ can be asymptotically expanded as
    \begin{equation}
        \hat{g}(\gamma) \sim \frac{1}{2}\left(\frac{1}{\gamma} + \frac{1}{4\gamma^2} + \mathcal O(\frac{1}{\gamma^3})\right).
        \label{eqn:g hat asymptotic expansion}
    \end{equation}
    Thus, $1/\gamma - \hat{g}(\gamma) \sim 1/2\,\Theta (1/\gamma)$, and the relative error $r$ of a single term in the CRLB expression is upper bounded by introducing a positive parameter $\epsilon$, which is
    \begin{equation}
        \begin{aligned}
        r & \overset{(a)}{\leq} \frac{\delta/\gamma^2}{\lvert \,\lvert 1/\gamma - \hat{g}(\gamma)\rvert - \lvert \hat{g}(\gamma) -  g(\gamma)\rvert\,\rvert} \\
        & \overset{(b)}{\leq} \frac{\delta/\gamma^2}{1/((2+\epsilon)\gamma) - \delta/\gamma^2} \\
        & = \frac{\delta}{\gamma/(2+\epsilon) - \delta}, \\
        \end{aligned}
        \label{eqn:upper bound of r}
    \end{equation}
    where (a) comes from applying the triangle inequality, and (b) comes from assuming sufficiently large $\gamma$ such that the denominator is positive (note that this can be done due to~\eqref{eqn:g hat asymptotic expansion}).
    In order to let this upper bound hold, it should be satisfied that $\gamma > (2+\epsilon)\delta$, and the parameter $\epsilon$ should satisfy 
    \begin{equation}
        \frac{1}{\gamma} - \hat{g}(\gamma) > \frac{1}{(2+\epsilon)\gamma}, \quad\forall \gamma > (2+\epsilon)\delta,
    \end{equation}
    which is equivalent to 
    \begin{equation}
        \epsilon > \frac{1}{1-\gamma \hat{g}(\gamma)}-2,\quad \forall \gamma > (2+\epsilon)\delta.
        \label{eqn:ineq for epsilon}
    \end{equation}
    Such $\epsilon>0$ exists. Since the function $ 1/(1-\gamma \hat{g}(\gamma))$ is decreasing for sufficiently large $\gamma$, the inequality~\eqref{eqn:ineq for epsilon} is equivalent to 
    \begin{equation}
        \epsilon > \frac{1}{1-(2+\epsilon)\delta\, \hat{g}((2+\epsilon)\delta)}-2.
    \end{equation}
    In fact, choosing $\epsilon=4$ will satisfy all the conditions above when $\delta = 0.07$. Thus, from~\eqref{eqn:upper bound of r}, the relative error is upper bounded by 
    \begin{equation}
        r \leq \frac{0.07}{\gamma/6 - 0.07},\quad \forall \gamma > 0.42.
    \end{equation}
    We can easily see from the above inequality that this approximation becomes arbitrarily good when $\gamma\to\infty$, and the decreasing rate is of order $\mathcal{O}(1/\gamma)$. This completes the proof.

\red{
\section{Proof of Lemma~\ref{lemma:wirtinger_der_vM-EM}} \label{app:proof_Wirtinger_derivative}
    {\it {Proof.}}
    Let $z=\exp(\ri \hat{\varphi})$ be the complex representation of the estimator $\hat\varphi$, and 
    \begin{equation}
        w = w(\hat{\varphi}, {\bm s}) = \frac{2\beta}{\sigma_v^2 \kappa_0}\sum_{\ell=0}^{L-1}\exp(-\ri \psi_\ell) s_\ell \frac{\mu_\ell}{|\mu_\ell|} 
    \end{equation}
    be the scaled intermediate result in the proposed vM-EM algorithm ({\bf Algorithm~\ref{alg:vM-EM}}, line 7). Then, on convergence, $\hat\varphi$ is  the fixed point of the algorithmic iteration, i.e., the complex number $w$ is parallel to $z$, which is equivalent to $w^*z\in\mathbb{R}$. 
    
    Define $p(z,w) = (zw^* - z^*w)/(2\ri): \mathbb{C}^2 \to \mathbb{R}$, then the convergence of this algorithm is equivalent to $w^*z\in\mathbb{R}$, which is further equivalent to $p(z,w)=0$. Since both $z=z(\hat{\varphi})$ and $w=w(\hat{\varphi}, {\bm s})$ are functions of the estimator $\hat{\varphi}$, the output of the vM-EM algorithm is a root of the equation $p=0$, i.e.,  
    \begin{equation}
        p(z(\hat{\varphi}),w(\hat{\varphi}, {\bm s}))=0 .
        \label{eqn:vM-EM_balance}
    \end{equation}
    Notice that if the observed signal is noiseless, i.e., $s_\ell = \mu_\ell^{(0)}:=\alpha+\beta\exp(\ri(\psi_\ell+\varphi))$, then the true value $\varphi$ is a solution to the equation~\eqref{eqn:vM-EM_balance}. This can be easily seen from the fact that if $\hat{\varphi} = \varphi$, then $\mu_\ell = \mu_\ell^{(0)}$.
    However, generally the input of the algorithm ${\bm s}$ is noisy, which is modeled by $s_\ell = |\mu_\ell^{(0)} + v_\ell|$, where $v_\ell \sim$ i.i.d. $\mathcal{CN}(0, \sigma_v^2)$ are the thermal noise at the $\ell$-th power sensor. Thus, the estimator $\hat\varphi$ can be viewed as a perturbed version of the true value $\varphi$, i.e., $\hat{\varphi} = \hat{\varphi}(\bm v)$, which is a function of the noise $\bm v$. 
    Thus, we aim to find the {\it Wirtinger derivative}~\cite{remmert1991theory} $\nabla_{\bm v}(\hat{\varphi}({\bm v}))$, where the gradient operator $\nabla_{{\bm v}}$ acting on $f: \mathbb{C}^L\to \mathbb{C}$ is defined through its components:
    \begin{equation}
        \begin{aligned}
        \nabla_{\bm v}f &= \frac{1}{2}\left( \nabla_{\Re({\bm v})} f -\ri \nabla_{\Im({\bm v})} f \right) \\
        &={\frac{1}{2}} \left( \frac{\partial f}{\partial v_{0, R}} -\ri \frac{\partial f}{\partial v_{0, I}}, \cdots,  \frac{\partial f}{\partial v_{L-1, R}} -\ri \frac{\partial f}{\partial v_{L-1, I}}\right)\T. 
        \end{aligned}
    \end{equation}
    Choose $\ell\in\{L\}$, we first evaluate $\partial \hat{\varphi}/\partial v_\ell$. Taking the derivative of both sides of~\eqref{eqn:vM-EM_balance} w.r.t. $v_\ell\in\mathbb{C}$ and using the derivative formula for implicit functions, we obtain 
    \begin{equation}
        \frac{\partial \hat\varphi}{\partial v_\ell}=-\frac{\frac{\partial p}{\partial w} \frac{\partial w}{\partial v_\ell}+\frac{\partial p}{\partial w^*} \frac{\partial w^*}{\partial v_\ell}}{\left(\frac{\partial p}{\partial z} \frac{\partial z}{\partial \hat\varphi}+\frac{\partial p}{\partial z^*} \frac{\partial z^*}{\partial \hat\varphi}\right)+\left(\frac{\partial p}{\partial w} \frac{\partial w}{\partial \hat\varphi}+\frac{\partial p}{\partial w^*} \frac{\partial w^*}{\partial \hat\varphi}\right)},
        \label{eqn:implicit_function_derivative} 
    \end{equation}
    where the {\it Wirtinger derivative} is applied to functions of complex variables, and the ordinary derivative is applied to functions of real variables. 
    By calculating the Wirtinger derivatives of $p$ w.r.t. $z$, $z^*$, $w$ and $w^*$, as well as $z$ w.r.t. $\hat{\varphi}$, we obtain 
    \begin{equation}
        \begin{aligned}
        \frac{\partial p}{\partial z}=\frac{\ri}{2} w^*, & \frac{\partial p}{\partial z^*}=-\frac{\ri}{2} w, \\
        \frac{\partial p}{\partial w}=-\frac{\ri}{2} z^*, & \frac{\partial p}{\partial w^*}=\frac{\ri}{2} z, \\
        \frac{\partial z}{\partial \hat\varphi}=\ri z, & \frac{\partial z^*}{\partial \hat\varphi}=-\ri z^*. 
        \end{aligned}
        \label{eqn:basic_derivatives}
    \end{equation}
    Substituting~\eqref{eqn:basic_derivatives} into~\eqref{eqn:implicit_function_derivative}, we obtain 
    \begin{equation}
        \frac{\partial \hat\varphi}{\partial v_\ell} = -\frac{\ri}{2}\frac{\left( z\frac{\partial w^*}{\partial v_\ell} -z^* \frac{\partial w}{\partial v_\ell} \right)}{-\Re(z^*w) +\Im(z^* \frac{\partial w}{\partial \hat{\varphi}})}. 
    \end{equation}
    Let $\frac{\partial \varphi}{\partial v_\ell} = -N_\ell / D_\ell$, where $N_\ell$ and $D_\ell$ are defined to be the numerator and denominator of the above equation respectively. In order to obtain expressions for $N_\ell$ and $D_\ell$, we first evaluate 
    \begin{equation}
        \begin{aligned}
            \frac{\partial w}{\partial v_\ell}&=\frac{2\beta}{\sigma_v^2 \kappa_0} \sum_{\ell'=0}^{L-1} e^{-\ri \psi_{\ell'}} \frac{\mu_{\ell'}}{|\mu_{\ell'}|} \frac{\partial s_{\ell'}}{\partial v_\ell} \\
            &= \frac{2\beta}{\sigma_v^2 \kappa_0} e^{-\ri \psi_\ell}\frac{\mu_\ell}{|\mu_\ell|} \frac{(\mu_\ell^{(0)} + v_\ell)^*}{2|\mu_\ell^{(0)} + v_\ell|}, 
        \end{aligned}
    \end{equation}
    where the formula $\frac{\partial |\rho|}{\partial \rho}  = \rho^*/(2|\rho|)$ is used. 
    Thus, we obtain 
    \begin{equation}
        \begin{aligned}
            N_\ell &= \frac{\ri}{2} \left( z\frac{\partial w^*}{\partial v_\ell} -z^* \frac{\partial w}{\partial v_\ell} \right) \\
            &= \frac{2\beta}{\sigma_v^2 \kappa_0} \frac{(\mu_\ell^{(0)} + v_\ell)^*}{2s_\ell |\mu_\ell|} \left(-\Im\{e^{\ri \theta_\ell} \mu_\ell^* \}\right) \\
            &= -\frac{2\alpha\beta}{\sigma_v^2 \kappa_0} \frac{(\mu_\ell^{(0)} + v_\ell)^*}{2s_\ell |\mu_\ell|} \sin(\theta_\ell). 
        \end{aligned}
    \end{equation}
    
    Similarly, by evaluating $\partial w/\partial \hat{\varphi}$ as 
    \begin{equation}
        \begin{aligned}
        \frac{\partial w}{\partial \hat{\varphi}} &= \frac{2\beta}{\sigma_v^2 \kappa_0} \sum_{\ell=0}^{L-1} e^{-\ri \psi_\ell} s_\ell \frac{\partial}{\partial \hat{\varphi}}\left(\frac{\mu_\ell}{|\mu_\ell|}\right) \\
        &= \frac{2\beta}{\sigma_v^2 \kappa_0} \sum_{\ell=0}^{L-1} e^{-\ri \psi_\ell} s_\ell  \frac{\mu_\ell}{|\mu_\ell|^3} {\ri} \Im \left\{\mu_\ell^* \frac{\partial \mu_\ell}{\partial \hat{\varphi}}  \right\} \\
        &=  \frac{2\beta}{\sigma_v^2 \kappa_0} \sum_{\ell=0}^{L-1} e^{-\ri \psi_\ell} s_\ell  \frac{\mu_\ell}{|\mu_\ell|^3} {\ri} \beta(\beta + \alpha\cos(\theta_\ell)), 
        \end{aligned}
    \end{equation}
    we get the denominator $D=D_\ell$ as 
    \begin{equation}
        \begin{aligned}
        D &= -\Re(z^*w) +\Im(z^* \frac{\partial w}{\partial \hat{\varphi}}) \\
        &= -\Re\{z^*(w+\ri \frac{\partial w}{\partial \hat{\varphi}})\} \\
        &= -\frac{2\beta}{\sigma_v^2 \kappa_0} \Re\left\{ \sum_{\ell=0}^{L-1}e^{-\ri \theta_\ell}  \frac{s_\ell\mu_\ell}{|\mu_\ell|}\left( 1-\beta\cdot \frac{\beta+\alpha\cos(\theta_\ell)}{|\mu_\ell|^2} \right) \right\} \\
        &= \frac{2\alpha\beta}{\sigma_v^2 \kappa_0} \langle {\bm s}, {\bm x} \rangle,
        \end{aligned}
    \end{equation}
    where ${\bm x} = (x_0, x_1, \cdots, x_{L-1})\T$, and its components defined as 
    \begin{equation}
        \begin{aligned}
            x_\ell &= \Re\{ e^{-\ri \theta_\ell} \frac{\mu_\ell}{|\mu_\ell|} \}(1- \beta(\beta+\alpha\cos(\theta_\ell))/|\mu_\ell|^2) \\
            &= \frac{(\beta+\alpha\cos(\theta_\ell))(\alpha+\beta\cos(\theta_\ell))}{|\mu_\ell|^3}. 
        \end{aligned}
    \end{equation}
    Finally, the squared norm of the gradient $\nabla_{\bm v}\hat{\varphi}$ is given by 
    \begin{equation}
        \begin{aligned}
        \|\nabla_{\bm v}\hat{\varphi}\|^2 &= \sum_{\ell=0}^{L-1}\left| \frac{N_\ell}{D} \right|^2 \\
        &= \frac{1}{|\langle {\bm s}, {\bm x}\rangle|^2}\sum_{\ell=0}^{L-1}  \frac{\sin^2(\theta_\ell)}{|\mu_\ell|^2},  
        \end{aligned}
    \end{equation}
    which completes the proof. 

\section{Proof of Lemma~\ref{lemma:ODE-bound}} \label{app:proof_ODE_bound}
From the differential representation of the estimation error~\eqref{eqn:differential_of_error}, for all $t\in[0,1]$, we obtain 
\begin{equation}
    {\rm d}|\Delta\varphi| \leq 2\|\nabla_{t{\bm v}}\hat{\varphi}\|\cdot \|{\bm v}\|{\rm d}t.
\end{equation}
Since $H_L = H_\infty + \mathcal{O}(L^{-1})$, we can conclude that the upper limit and the lower limit of the sequence $H_L$ converges to the same limit $H_\infty$ as $L\to\infty$, and this convergence is uniform in $\varphi\in[0,2\pi]$. Specifically, if we denote 
\begin{equation}
    \begin{aligned}
    \underline{H}_L &:= \inf_{k\geq L, \varphi\in[0,2\pi]}{H_k}, \\
    \overline{H}_L &:= \sup_{k\geq L, \varphi\in[0,2\pi]}{H_k}, 
    \end{aligned}
\end{equation}
then $\underline{H}_L = H_\infty+\mathcal{O}(L^{-1})$, and $\overline{H}_L = H_\infty+\mathcal{O}(L^{-1})$. The same definitions and upper/lower limiting properties hold for the sequences $G_L$ and $X_L := \|{\bm x}\|/\sqrt{L}$. It follows immediately that 
\begin{equation}
    \begin{aligned}
        & \underline{H}_L \leq H_L \leq \overline{H}_L, \\
        & \underline{G}_L \leq G_L \leq \overline{G}_L, \\
        & \underline{X}_L \leq X_L \leq \overline{X}_L. 
    \end{aligned}
\end{equation}

Thus, by~{\bf Lemma~\ref{lemma:wirtinger_der_vM-EM}}, we can further bound the increasing rate of $|\Delta\varphi|$ from above, i.e., 
\begin{equation}
    {\rm d}|\Delta\varphi| \leq \frac{2\sqrt{L\overline{H}_L}}{|\langle{\bm s}(t), {\bm x}\rangle|} \cdot \|{\bm v}\|{\rm d}t
    \label{ineq:diff_of_error}
\end{equation}
where ${\bm s}(t) = |{\bm \mu}^{(0)} + t{\bm v}|$, and the components of ${\bm \mu}^{(0)}\in\mathbb{C}^L$ is defined as $\mu_\ell^{(0)}:= \alpha+\beta\exp(\ri (\psi_\ell + \varphi))$. In order to obtain an upper bound for the increasing rate of $|\Delta\varphi|$ w.r.t. $t$, we first try to lower-bound the inner product $\langle{\bm s}(t), {\bm x}\rangle$. $\forall t\in[0,1]$: 
\begin{equation}
    \begin{aligned}
    \langle{\bm s}(t), {\bm x}\rangle &= \langle{\bm s}(t) - |{\bm \mu}|, {\bm x}\rangle + \langle|{\bm\mu}|,{\bm x}\rangle\\
    &= \langle{\bm s}(t) - |{\bm \mu}|, {\bm x}\rangle + LG_L \\
    &\geq L\underline{G}_L-\|{\bm s}(t) - |{\bm \mu}|\|\cdot\|{\bm x}\| \\
    & \geq L\underline{G}_L-(\|{\bm s}(t) - |{\bm \mu}^{(0)}|\|+\||{\bm \mu}^{(0)}| - |{\bm \mu}|\|)\cdot\|{\bm x}\| \\
    &\overset{(a)}{\geq} L\underline{G}_L-(t\|{\bm v}\| + \sqrt{L}\beta|\Delta\varphi|)\cdot\|{\bm x}\| \\
    &\geq L\left[\underline{G}_L-(\frac{t\|{\bm v}\|}{\sqrt{L}}+\beta|\Delta\varphi|)\cdot \overline{X}_L \right]\\
    \end{aligned}
\end{equation}
where (a) comes from applying the triangle inequality to the definition of ${\bm s}(t)$ and ${\bm \mu}^{(0)}$. Combining the above inequality with the differential inequality~\eqref{ineq:diff_of_error}, we obtain 
\begin{equation}
    {\rm d}|\Delta\varphi|\leq \frac{2\sqrt{\overline{H}_L}}{\underline{G}_L-(\frac{t\|{\bm v}\|}{\sqrt{L}}+\beta|\Delta\varphi|)\cdot \overline{X}_L }\cdot \frac{\|{\bm v}\|}{\sqrt{L}} {\rm d}t,
    \label{ineq:increasing-rate-constraint}
\end{equation}
where the initial condition is $|\Delta\varphi|(t=0) = |\Delta\varphi|({\bm v} = 0)=0$. In order to solve this differential inequality, we construct an ordinary differential equation (ODE) with zero initial condition as 
\begin{equation}
    \frac{\d y}{\d u} = g_L(y, u) := \frac{2\sqrt{\overline{H}_L}}{\underline{G}_L - \overline{X}_L\left(u+\beta y\right)}, ~~~~y(0)=0.  \label{eqn:ODE}
\end{equation}
Since $u=0$ is a non-singular point of the ODE, there exists some small positive number $\delta=\delta(L)>0$ such that the solution to~\eqref{eqn:ODE} uniquely exists in some closed interval $I_{\delta} = [0,\delta]$~\cite{ince1956ordinary}. Note that this $\delta$ is chosen for some given $L$, but the existence of the ODE solution, as well as all the following properties, hold true for arbitrary larger values of $L$ in the same interval $I_\delta$, because of the monotone property of the sequences $\overline{H}_L$, $\underline{G}_L$, and $\overline{X}_L$. 

Furthermore, there exists some positive constant $C>0$ such that $\forall u\in I_{\delta}$, $y(u)\leq Cu$. Particularly, since $g_L(y,u)$ is an increasing function of $y$ and $u$, $C$ can be chosen to be 
\begin{equation}
    C(\delta, L) = \frac{2\sqrt{\overline{H}_L}}{\underline{G}_L-\overline{X}_L (\delta+\beta y(\delta))}. 
\end{equation} 
Since the solution $y(u)$ of this ODE~\eqref{eqn:ODE} characterizes the maximum value attainable for arbitrary smooth function subject to the increasing rate constraint~\eqref{ineq:increasing-rate-constraint}, the solution $y(u)$ in $I_\delta$ serves as an upper bound to the unknown function $|\Delta\varphi|(t{\bm v})$ as a function of $t$, i.e., 
\begin{equation}
    |\Delta\varphi|(t{\bm v}) \leq y\left(t\frac{\|{\bm v}\|}{\sqrt{L}}\right), \quad \forall t\frac{\|{\bm v}\|}{\sqrt{L}} \in I_\delta.  
\end{equation}
Thus, by further applying $y(u)\leq Cu$, we can infer that $\forall t: 0\leq t \leq \delta (\|{\bm v}\|/\sqrt{L})^{-1}$,
\begin{equation}
    |\Delta\varphi|(t{\bm v}) \leq C t\frac{\|{\bm v}\|}{\sqrt{L}}.
\end{equation} 
If we apply the constraint $\|{\bm v}\|/\sqrt{L}\leq \delta$, then $t$ can be chosen to $t=1$, which completes the proof. 
}

\section*{Acknowledgment}
We would like to thank Dr. Yajun Zhao from ZTE Corporation for his helpful discussions and constructive suggestions on Sensing RIS, and thank Dr. Qian Ma from Southeast University for his professional advice on sensing metasurfaces.   

\footnotesize
\bibliographystyle{IEEEtran}
\bibliography{SensingRIS, IEEEabrv}

\vspace{-1.5cm}
\begin{IEEEbiographynophoto}{Jieao Zhu}
    received the B.E. degree in electronic engineering, and the second B.S. degree in applied mathematics from Tsinghua University in 2021. Currently, he is pursuing the Ph.D. degree in Department of Electronic Engineering at Tsinghua University, Beijing, China. 
    His research interests include electromagnetic information theory (EIT), coding theory, and quantum computing. 
    He has received the National Scholarship in 2018 and 2020, and the Excellent Graduates of Beijing in 2021.   
\end{IEEEbiographynophoto}

\vspace{-1cm}
\begin{IEEEbiographynophoto}{Kunzan Liu}
    received the B.E. degree in electronic engineering from Tsinghua University in 2022. He is currently pursuing the Ph.D. degree at Massachusetts Institute of Technology. His research interests lie in the fields of dimensionality reduction for wireless communications, and multiphoton microscopy and computational imaging for biological understandings.  
    He has received National Scholarship from 2019 to 2021, and Excellent Bachelor Dissertation of Beijing in 2022. 
\end{IEEEbiographynophoto}

\vspace{-1cm}
\begin{IEEEbiographynophoto}{Zhongzhichao Wan}
    received the B.E. degree in electronic engineering. He is currently pursuing the Ph.D. degree in Department of Electronic Engineering at Tsinghua University, Beijing, China. His research interests include electromagnetic information theory (EIT), coding theory, and channel modeling.   
    He has received the Excellent Graduates of Tsinghua University in 2020. 
\end{IEEEbiographynophoto}

\vspace{-1cm}
\begin{IEEEbiographynophoto}{Linglong Dai}
    (M’11-SM’14-F’21)  received the B.S. degree from Zhejiang University, Hangzhou, China, in 2003, the M.S. degree from the China Academy of Telecommunications Technology, Beijing, China, in 2006, and the Ph.D. degree from Tsinghua University, Beijing, in 2011. From 2011 to 2013, he was a Post-Doctoral Researcher with the Department of Electronic Engineering, Tsinghua University, where he was an Assistant Professor from 2013 to 2016, an Associate Professor from 2016 to 2022, and has been a Professor since 2023. His current research interests include massive MIMO, reconfigurable intelligent surface (RIS), millimeter-wave and Terahertz communications, wireless AI, and electromagnetic information theory. He has received the National Natural Science Foundation of China for Outstanding Young Scholars in 2017, the IEEE ComSoc Leonard G. Abraham Prize in 2020, the IEEE ComSoc Stephen O. Rice Prize in 2022, and the IEEE ICC Best Demo Award in 2022. He was elevated as an IEEE Fellow in 2021.
\end{IEEEbiographynophoto}

\vspace{-1cm}
\begin{IEEEbiographynophoto}{Tie Jun Cui}
    (M’98–SM’00–F’15) received the B.Sc., M.Sc., and Ph.D. degrees in electrical engineering from Xidian University, Xi’an, China, in 1987, 1990, and 1993, respectively. In March 1993, he joined the Department of Electromagnetic Engineering, Xidian University, and was promoted to an Associate Professor in November 1993. In July 1997, he joined the Center for Computational Electromagnetics, Department of Electrical and Computer Engineering, University of Illinois at Urbana-Champaign, first as a Postdoctoral Research Associate and then a Research Scientist. In September 2001, he was a Cheung-Kong Professor with the Department of Radio Engineering, Southeast University, Nanjing, China. Currently he is the Chief Professor of Southeast University, and the Director of State Key Laboratory of Millimeter Waves. 
    Dr. Cui’s research interests include metamaterials and computational electromagnetics. Dr. Cui received the Natural Science Award (first class) from the Ministry of Education, China, in 2011, and the National Natural Science Awards of China (second class, twice), in 2014 and 2018, respectively. 
    Dr. Cui is the Academician of Chinese Academy of Science, and IEEE Fellow.  In 2019-2022, he was ranked in the top 1\% for the highly cited papers in the field of Physics by Clarivate Web of Science (Highly Cited Researcher).
\end{IEEEbiographynophoto}

\begin{IEEEbiographynophoto}{H. Vincent Poor}
(S’72-M’77-SM’82-F’87) received the Ph.D. degree in EECS from
Princeton University in 1977. From 1977 until 1990, he was on the faculty of the
University of Illinois at Urbana-Champaign. Since 1990 he has been on the faculty at Princeton, where he is currently the Michael Henry Strater University Professor. During 2006 to 2016, he served as the dean of Princeton’s School of Engineering and Applied Science. He has also held visiting appointments at several other universities, including most recently at Berkeley and Cambridge. His research interests are in the areas of information theory, machine learning and network science, and their applications in wireless networks, energy systems and related fields. Among his publications in these areas is the recent book Machine Learning and Wireless Communications. (Cambridge University Press, 2022). Dr. Poor is a member of the National Academy of Engineering and the National Academy of Sciences and is a foreign member of the Chinese Academy of Sciences, the Royal Society, and other national and international academies. He received the IEEE Alexander Graham Bell Medal in 2017.
\end{IEEEbiographynophoto}

\end{document}